\documentclass[natbib,letterpaper,iop,twocolumn]{emulateapj}
\usepackage{amstext}
\usepackage{amsmath}
\usepackage{amssymb}
\usepackage[colorlinks=true,linkcolor=blue,citecolor=blue]{hyperref}
\begin{document}

\newcommand{\beq}{\begin{equation}}
\newcommand{\eeq}{\end{equation}}
\newcommand{\beqar}{\begin{eqnarray}}
\newcommand{\eeqar}{\end{eqnarray}}
\newcommand{\e}{\varepsilon}
\newcommand{\rt}{r_{\rm t}}
\newcommand{\rs}{r_{\rm s}}
\newcommand{\Mbh}{M_{\rm bh}}
\newcommand{\rp}{r_{\rm p}}
\newcommand{\risco}{r_{\rm isco}}
\newcommand{\rb}{r_{\rm b}}
\newcommand{\Ra}{R_{\rm a}}
\newcommand{\Ma}{M_{\rm a}}
\newcommand{\vinf}{v_\infty}
\newcommand{\rhoinf}{\rho_\infty}
\newcommand{\mach}{\mathcal M_\infty}
\newcommand{\mhl}{\dot M_{\rm HL}}
\newcommand{\lhl}{\dot L_{\rm HL}}
\newcommand{\Hrho}{H_\rho}
\newcommand{\Rs}{R_{\rm s}}
\newcommand{\erho}{\epsilon_\rho}
\newcommand{\cs}{c_{\rm s,\infty}}

\title{Asymmetric Accretion Flows within a Common Envelope}

\author{Morgan MacLeod and Enrico Ramirez-Ruiz}
\affil{Department of Astronomy and
  Astrophysics, University of California, Santa Cruz, CA
  95064}
  
\begin{abstract} 
This paper examines flows in the immediate vicinity of stars and compact objects dynamically inspiralling within a common envelope (CE). 
Flow in the vicinity of the embedded object is gravitationally focused leading to drag and potential to gas accretion. 
This process has been studied numerically and analytically in the context of  Hoyle-Lyttleton accretion (HLA).  
Yet, within a CE, accretion structures may span a large fraction of the envelope radius, and in so doing sweep across a substantial radial gradient of density. 
We quantify these gradients using detailed stellar evolution models for a range of CE encounters.  We provide estimates of typical scales in CE encounters that involve main sequence stars, white dwarfs, neutron stars, and black holes with giant-branch companions of a wide range of masses.  We apply these typical scales to hydrodynamic simulations of 3D HLA with an upstream density gradient.  
This density gradient breaks the symmetry that defines HLA flow, and imposes an angular momentum barrier to accretion. Material that is focused into the vicinity of the embedded object thus may not be able to accrete. 
As a result, accretion rates drop dramatically, by 1-2 orders of magnitude, while drag rates are only mildly affected. 
We  provide fitting formulae to the numerically-derived rates of drag and accretion as a function of the density gradient. 
The reduced ratio of accretion to drag  suggests that objects that can efficiently gain mass during CE evolution, such as black holes and neutron stars, may grow  less than implied by the HLA formalism. 
\end{abstract}

\maketitle

\section{Introduction}
As stars evolve off of the main sequence their radius grows dramatically. This expansion has major consequences for stars in close binaries. A common envelope (CE) phase occurs when one star grows to the point that it engulfs its more compact companion within a shared envelope \citep{Paczynski:1976uo}. This process is not rare \citep{Kochanek:2014uv}; the majority of stars exist in binary or multiple systems \citep[e.g.][]{Duchene:2013il}, and interactions or mergers should mark the evolution of $\gtrsim 30$\% of massive stars \citep{Sana:2012gu,deMink:2014fl}. During the CE phase, orbital energy and angular momentum are shared with the surrounding gaseous envelope tightening the orbit of the embedded binary cores \citep{Paczynski:1976uo,Iben:1993ka,Taam:2000jy,Ivanova:2013co}. Whether the surrounding envelope is ejected, and the binary survives the CE phase, depends on the efficiency with which energy \citep{Webbink:1984jd,Iben:1993ka,Han:1995ve,Tauris:2001gj} and angular momentum \citep{Nelemans:2000wn,vanderSluys:2006ht,Toonen:2013dv} can be harnessed to expel envelope material. 

CE episodes, and their outcomes, are therefore critical in shaping populations of close binaries \citep[e.g.][]{Belczynski:2002gi,Stairs:2004go,Kalogera:2007kh,Toonen:2013dv} as well as their merger products \citep{Sana:2012gu,deMink:2013js,deMink:2014fl}. Substantial progress has been made by constraining the efficiencies of CE ejection compared to the change in orbital energy via the parameter $\alpha_{\rm CE}$ \citep{Webbink:1984jd,Iben:1993ka}, or the change in orbital angular momentum with the parameter $\gamma_{\rm CE}$ \citep{Nelemans:2000wn}. These binary population synthesis methods are powerful but also very sensitive to the details of stellar envelope structure  at the onset of the CE event \citep[e.g.][]{Han:1995ve,Tauris:2001gj,DeMarco:2011cr,Soker:2013jj}, as well as the specific post-CE binary system under consideration; there do not appear to be universal answers with respect to the outcome of CE events \citep{Ivanova:2013co}.

Theoretical efforts to constrain the physics and properties of CE dispersal have recently been extended to include three-dimensional hydrodynamic simulations of the early inspiral process \citep{Livio:1988io,Terman:1994er,Taam:1994km,Terman:1995cy,Terman:1996gs,Sandquist:1998kk,Ricker:2007kc,Passy:2012jo,Ricker:2012gu}. By necessity, the embedded object in these simulations is described only by its gravitational influence on the gas,
 and the region that would represent the companion may be unresolved  \citep[e.g.][]{Passy:2012jo}, or resolved by a few numerical cells \citep[e.g.][]{Ricker:2012gu}.
Determining flow properties in the immediate vicinity of an embedded object remains an unmet challenge because of the huge range of spatial and temporal scales described by a CE episode \citep{Taam:2010dc}. 

The traditional approach to understanding flows around an embedded object during CE evolution has focused on analytic work by \citet{Hoyle:1939fl}, \citet{Bondi:1944ty}, and \citet{Bondi:1952wx}. These studies developed an analytic understanding of the nature of accretion onto a gravitational source moving supersonically through its surrounding medium \citep[see][for a recent review]{Edgar:2004ip}. 
Semi-analytical work has used these prescriptions to estimate the inspiral and accretion experienced by an object embedded in an envelope. In some cases, the co-evolution of the envelope and accretor are treated in 1D \citep[e.g.][]{Taam:1978dk,Meyer:1979wo,Taam:1978dk,Shankar:1991ju,Kato:1991ez,Siess:1999iu,Siess:1999im,Passy:2012ir}, while in others, the initial envelope properties are used as motivation for the relevant parameters for the embedded object \citep[e.g.][]{Livio:1988io,Chevalier:1993by,Iben:1993ka,Bethe:1998jv,Metzger:2012hg}. An issue, of course, is that stellar envelopes are not uniform density media, an effect that can be seen particularly clearly in recent 3D simulations by \citet{Passy:2012jo} and \citet{Ricker:2007kc,Ricker:2012gu}.

In this paper, we adopt an idealized approach to perform numerical simulations that explore the behavior of flows in the immediate vicinity of the embedded object.  In particular, accretion structures, as we discuss in  Section \ref{sec:scales}, typically span a large portion of the stellar radius \citep[e.g.][]{Iben:1993ka}. They impinge on a large portion of the envelope, and sweep across a huge radial gradient of envelope density \citep{Ricker:2012gu}. We extend traditional simulations of three-dimensional Hoyle-Lyttleton accretion (HLA) to consider the effects of these substantial density gradients on flow patterns, drag force on the object's motion, and the rate of mass and angular momentum accretion by the embedded object. 

This work builds on analytic considerations \citep{Dodd:1952uc,Illarionov:1975ul,Shapiro:1976gm,Davies:1980vt} and simulations of  non-axisymmetric flow conditions in HLA  \citep{Soker:1984tn,Soker:1986tg,Livio:1986vr,Fryxell:1988kb,Ruffert:1999tq}. However, since much of this work was motivated by inhomogeneities in wind-capture binaries, typical density gradients tend to be much milder than those experienced by an embedded object within a CE\footnote{For example, 3D calculations involving mild density gradients in HLA are currently being performed by  \citet{Raymer:2014wl}. }. We will compare our results in detail to conceptually similar but 2D simulations by \citet{Armitage:2000eg} to discuss the significant effect of 3D flow geometries.  
Of course, this idealized approach carries many compromises in that local simulations do not capture the full geometry, gravitational potential, eccentric orbital motions \citep{Passy:2012jo}, microphysics of energy sources or sinks \citep{Iben:1993ka,Ivanova:2013co}, or  disturbed background flow present in true CE  events \citep{Ricker:2012gu}. 
Even so, we will argue that the effects of a density gradient on HLA-like flows are so dramatic that a detailed understanding of these idealized cases carries important implications for CE evolution.

The remainder of this paper is organized as follows. In Section \ref{sec:scales}, we parameterize typical flow characteristics around an object embedded in a CE. Typical properties range from one to several density scale heights per characteristic accretion radius. In Section \ref{sec:methods} and \ref{sec:results} we describe the methods and results, respectively, of 3D hydrodynamic simulations of planar flow with an  imposed density gradient past a gravitating object. We derive fitting formulae for the drag force and for rates of accretion of mass and angular momentum as a function of density gradient in these simulations.  In Section \ref{sec:discussion}, we discuss the implications of our findings for CE events, and in Section \ref{sec:conclusion} we conclude.

\section{Stellar Properties, Typical Scales and Gradients}\label{sec:scales}

At the onset of a CE episode, there is a dynamical phase  following the loss of corotation between the envelope and the binary in which one object becomes embedded within the envelope of its companion and begins to spiral inward \citep{Podsiadlowski:2001tn}. 
In our simplified analysis, we will suppose that the properties of the companion's envelope are unperturbed by the presence of the embedded object. We note that this assumption is most justified when the embedded object is a small fraction of the total mass \citep{Iben:1993ka}, but we defer the reader to Section \ref{sec:disturb} for a discussion of the potential implications of a perturbed envelope. 
In this section, we analyze the characteristic dimensionless scales that parameterize the flows that arise during this phase of dynamical flow.

\subsection{Characteristic scales}

The first characteristic scale of the upstream flow is the Mach number, 
\beq
\mach = {\vinf \over \cs},
\eeq
where $\cs$ is the sound speed and $\vinf$ is the relative speed of motion through the gas. In the context of a stellar envelope, the relevant sound speed is that evaluated at the radius of the embedded object within the star. The speed of the incoming flow is approximately the circular velocity of the orbiting stellar cores at that radius,
$
\vinf = v_{\rm circ} = G \left( m_*(a) + M \right) / a
$
where $m_*(a)$ is the enclosed stellar mass at a given separation $a$, and is some fraction of the total stellar mass $M_*$.  To be embedded, $a$ must be less than the envelope radius, $R_*$. As we will show in the next subsection, typical Mach numbers are mildly supersonic, $\mach \sim 1.5-5$. 

The theory of gravitational accretion onto a supersonically moving point mass was first elaborated by \citet{Hoyle:1939fl}. It is useful here to frame the accretion flows that are set up within a stellar envelope in the context of those realized in HLA. The accretion radius, $\Ra$, is a gravitational cross section for material in supersonic motion to be focused toward the accretor,
\beq
\Ra  = \frac{2 G M}{\vinf^2}.
\eeq
This scale defines the material whose kinetic energy is sufficiently small that it will be focused to a line of symmetry downstream of the massive object and accrete \citep{Blondin:2012ep}. The accretion radius defines the material with which the embedded object can be expected to interact. 

The accretion radius and the properties of the upstream flow suggest a characteristic accretion rate onto the embedded object. The HLA accretion rate is defined by the flux of material with impact parameter less than the accretion radius,  
\beq\label{eq:mhl}
\mhl = \pi \Ra^2 \rho_\infty \vinf,
\eeq
where $\rhoinf$ is the density of the upstream flow.

\begin{figure}[tbp]
\begin{center}
\includegraphics[width=0.47\textwidth]{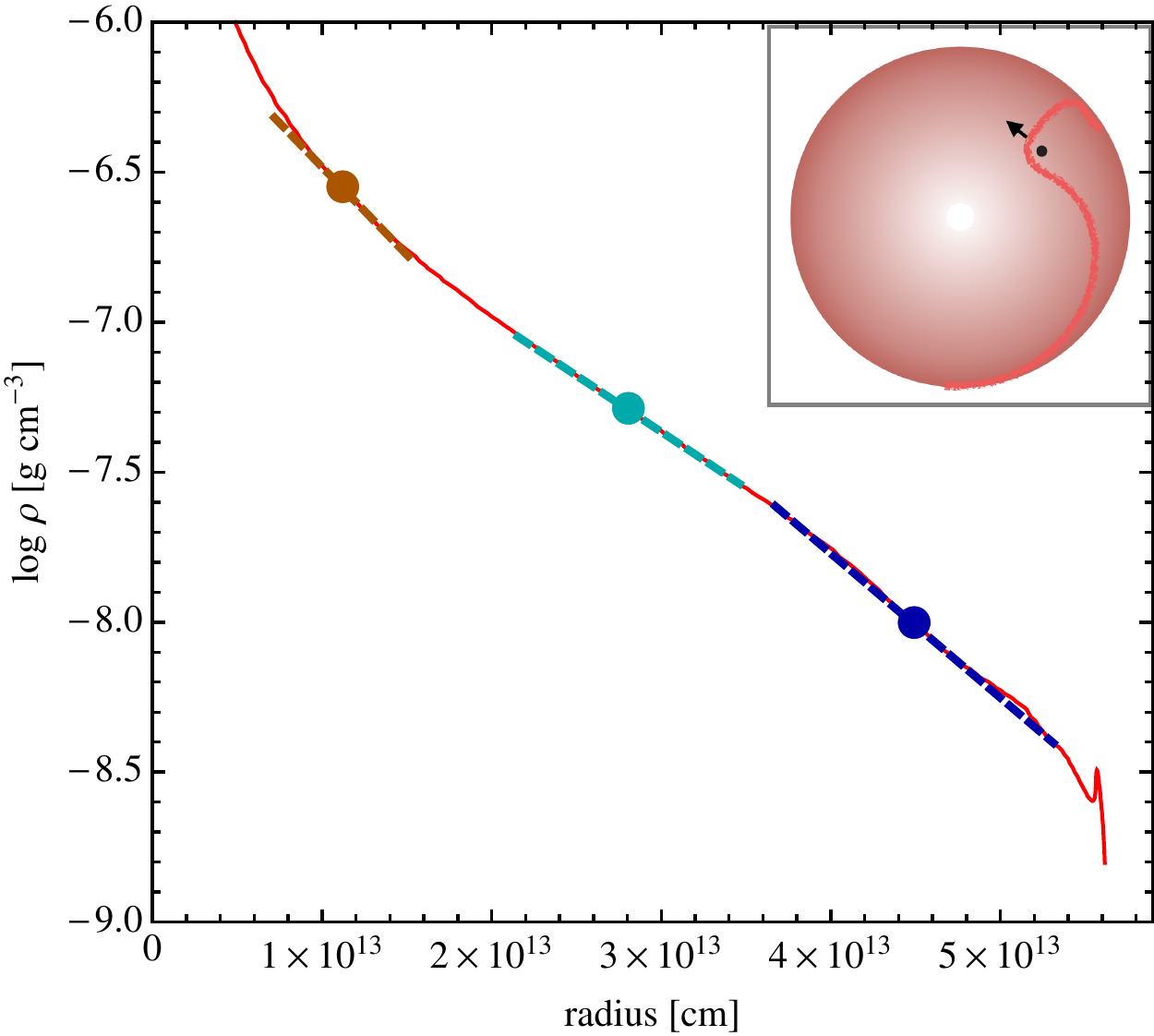}
\caption{Density gradients encountered by objects embedded within the envelope of a giant star (as drawn schematically in the inset image). In the inset cartoon, an embedded star orbits within its companions envelope, shocking the surrounding material.    In the main panel, we show the radial density profile of a $16M_\odot$ red supergiant (RSG), and assume the embedded object is a $1.4M_\odot$ neutron star (NS). The overplotted, dashed, lines show local exponential fits to the density profile following equation \eqref{reconstructeddens}. The dashed lines extend for $\pm \Ra$ as evaluated at the position of the embedded object (shown with points). We show examples of objects embedded at $0.2R_*$, $0.5R_*$ and $0.8R_*$.  Because it is a significant fraction of the envelope radius, $\Ra$ subtends a substantial density gradient implying that density asymmetries play an important role in shaping flow morphologies in CE evolution.    }
\label{fig:densityfit}
\end{center}
\end{figure}

Within a stellar envelope, the radial density gradient of the star ensures that the accretor experiences an upstream density gradient of incoming material. 
The key difference that this work will emphasize is the role of this gradient in shaping flow structures around the embedded object. We will characterize these density gradients in terms of the local density scale height, $\Hrho$. The density scale height is defined as 
\beq
\Hrho = -\rho \frac{dr }{d \rho},
\eeq
and is evaluated at the radius of the embedded object within the stellar envelope.
It will be useful to define a parameter that describes the number of density scale heights swept by the accretion radius,
\beq\label{eq:erho}
\epsilon_\rho \equiv {\Ra \over \Hrho}.
\eeq
In this context,  traditional HLA flow is characterized by upstream homogeneity, $\erho = 0$. 
The density profile may then be expanded around a point, $r_0$, as 
\beq\label{reconstructeddens}
\rho  \approx \rho (r_0) \exp\left( \erho {r_0-r \over \Ra} \right).
\eeq
An example of this expansion is shown for the radial run of density in a stellar model in Figure \ref{fig:densityfit}. We show exponential approximations, equation \eqref{reconstructeddens}, of the local density profile that extend $\pm \Ra$ from embedded radii of $0.2$, $0.5$, and $0.8R_*$. 
These expansions show that an object embedded at those radii encounters significant density inhomogeneity, which can be expected to play a key role in shaping the flow. 

\subsection{MESA Simulations}

We perform a set of stellar evolution calculations to explore the time evolution and range of typical values for the characteristic dimensionless scales outlined in the previous subsection. Our simulations use the MESA stellar evolution code, version 5527 \citep{Paxton:2011jf,Paxton:2013th}. We evolve stars of 1-16$M_\odot$ from the zero-age main sequence to their giant-branch expansion, during which a CE phase may be initiated. 

\subsubsection{Time Evolution}

\begin{figure}[tbp]
\begin{center}
\includegraphics[width=0.5\textwidth]{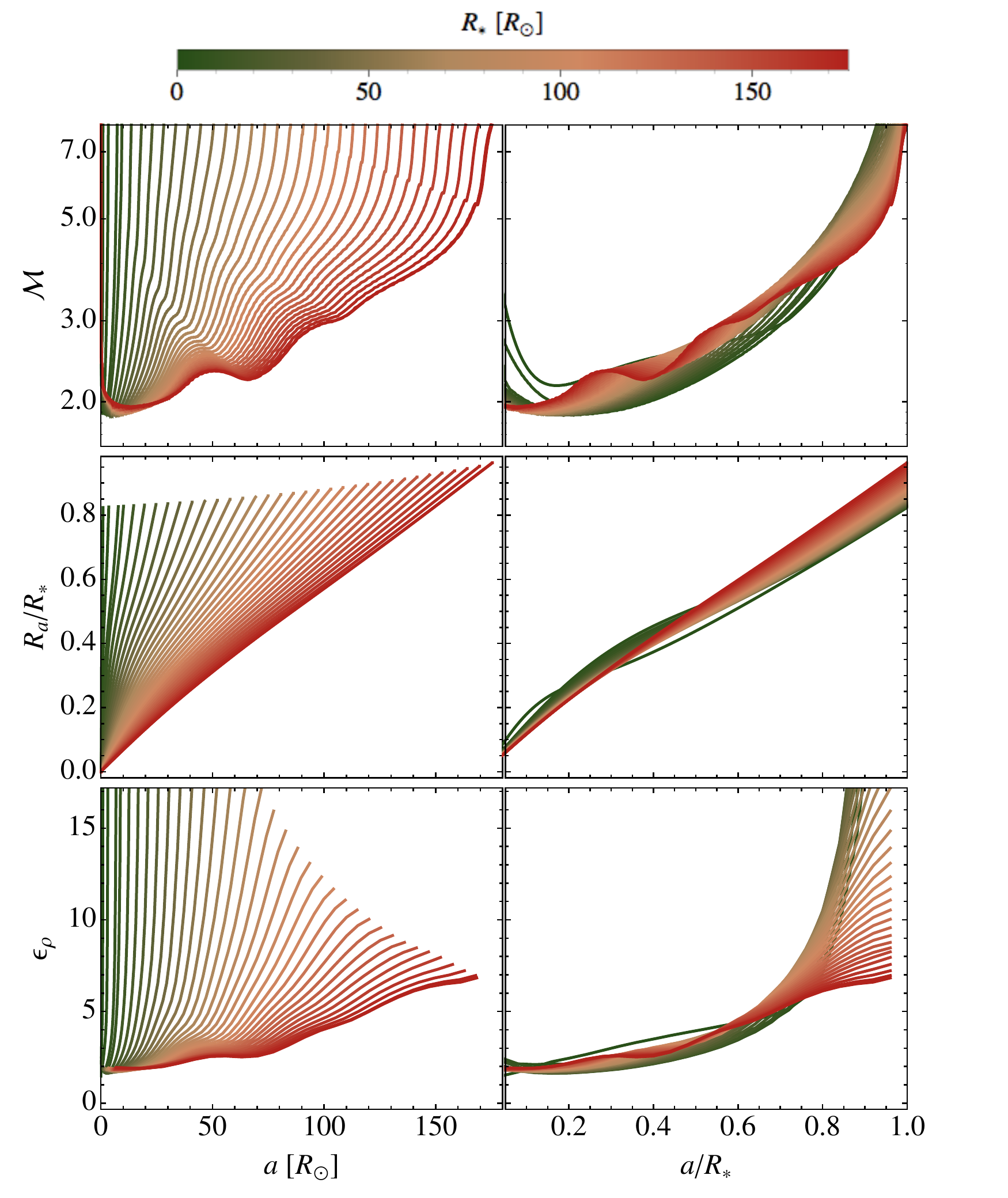}
\caption{Time evolution of characteristic flow parameters within the envelope of a $1M_\odot$ star as it evolves up the RGB, with the embedded object mass assumed to be  $0.7M_\odot$. The left series of figures show orbital separation in units of the solar radius, showing the expansion of the star as it ascends to the tip of the RGB and its maximum radius. The right-hand panels normalize the radius to the current stellar radius. From top to bottom, we plot Mach number, accretion radius as a fraction of stellar radius, and density gradient. Flow parameters vary with normalized radius, but are remarkably consistent over the RGB evolution.  This implies that regardless of where on the RGB the onset of CE occurs, we can expect relatively consistent hydrodynamic conditions. }
\label{fig:scalesEvolution}
\end{center}
\end{figure}

In Figure \ref{fig:scalesEvolution}, we show the time-evolution of a $1M_\odot$ star ascending the red giant branch (RGB). Under the  simplifying assumption that the envelope structure is not immediately perturbed by the the presence of the embedded object, we plot the upstream Mach number, the accretion radius as a fraction of the stellar radius, and the number of density scale heights per accretion radius, $\erho$. These panels display the conditions assuming the object is embedded at a separation $a$ from the center of its companion.  The left panels show separation in units of the solar radius, the right panels normalize the radius to a fraction of the stellar envelope radius $R_*$. 

The immediate conclusion we can draw from Figure \ref{fig:scalesEvolution} is that although the envelope expands dramatically during the course of the star's giant branch evolution, when normalized to a fraction of the radius, $a/R_*$, the basic upstream conditions for the flow are remarkably consistent. In Figure \ref{fig:scalesEvolution}, we color individual models based on their radius, which maps to the binary separation at the onset of CE. Typical Mach numbers range from $\mach \sim2$ in the deep interior to $\mach \gtrsim 5$ near the stellar limb. Although various local features develop due to opacity transitions and equation of state effects, this qualitative trend holds across the entire giant branch expansion. The accretion radius is a relatively constant fraction of the separation such that $\Ra / R_*$ is roughly proportional to $a/R_*$, as shown in the center-right panel of Figure \ref{fig:scalesEvolution}. The density gradient, as parameterized by $\erho$, is the parameter that exhibits the most temporal variation. In particular,  near the stellar limb, the density gradient  is extremely step, $\erho \gtrsim10$, in the early RGB phases, while maintaining more moderate values of $\erho \lesssim 8$ near the tip of the RGB.  
 As an object spirals to smaller $a/R_*$, the density gradient becomes shallower.

That these dimensionless scales remain relatively constant throughout the evolution implies that the global flow morphology in a wide variety of encounters can be sufficiently well described by a relatively small set of typical numbers. In the next subsection, we evaluate the conditions found in a representative set of CE encounters.

\subsubsection{Typical Scales in a Variety of Encounters}\label{sec:typicalencounters}

In this subsection we examine a small variety of representative CE events including those involving embedded planets, main sequence stars, and compact objects. 
In so doing, we hope to capture some of the diversity of CE events and explore how it manifests itself in the range of typical flow parameters. 

We provide estimates for the properties of the following systems:
\begin{itemize}

\item {\it Jupiter + $1M_\odot$ RGB star}: With increasing evidence that many low-mass stars host planetary systems -- including those with giant planets close to their host stars -- it is interesting to consider the final fate of these systems \citep[e.g.][]{Siess:1999iu,Siess:1999im,Sandquist:1998ik,Sandquist:2002in,Metzger:2012hg,Passy:2012ir}. A CE with extreme mass ratio may be initiated as the host star evolves up the giant branch. In these cases, the geometric cross section is very similar to the accretion radius, and the gravitational focusing of material from the upstream flow (as discussed in the remainder of this paper) is not a strong effect on the flow properties. \citet{Sandquist:2002in} has shown that more  relevant  effects include entrainment of planetary material into the star, and possible  consequences for the pollution of the host star. 

\item {\it $0.7 M_\odot$ white dwarf (WD) + $1M_\odot$ red giant branch (RGB) star:} This low-mass CE scenario might precede the formation of a double-WD binary, where the CE is initiated as the lower-mass star evolves off of the main sequence. \citet{Nelemans:2000wn} and \citet{vanderSluys:2006ht} model the formation of these double WD systems, with typical progenitor masses in the 1-3$M_\odot$ range. \citet{Hall:2013ej} have recently studied the structure and evolution of the low-mass stripped giant that would arise from such a system, and how this evolution would imprint itself during an ensuing planetary nebula phase. 

\item {\it Sun +  $2 M_\odot$ asymptotic giant branch (AGB) star:} In this case, a main sequence star of solar mass and radius is interacting with a CE donated by a slightly more massive AGB star. This scenario would lead to the formation of a close WD -- Main sequence binary. Broadly-defined, this main sequence with giant branch scenario is expected to be relatively common and was invoked by \citet{Paczynski:1976uo} in the original description of the CE scenario. 
 A cataclysmic variable state is among the possible outcomes from this channel if the post-CE binary is sufficiently close that it can be drawn into resumed mass transfer \citep{Paczynski:1976uo}. 
Variants of this scenario have been studied in the Double Core Evolution series of papers \citep{Taam:1989kl,Taam:1991dh,Yorke:1995et,Terman:1996gs}. A population synthesis of post-CE WD -- main sequence binaries has been recently undertaken by \citet{Toonen:2013dv}.

\item {\it Neutron star (NS) + 2, 8, 16 $M_\odot$ giants:} Following these scenarios in which a NS becomes embedded within the envelope of its companion, possible outcomes include close binaries consisting of a  NS and either a WD (in the lower mass companion cases) or He-star (in the higher companion-mass cases). The He-star could then undergo a core collapse supernova which might leave behind a double NS or NS-BH binary \citep{Postnov:2014fb}.  The hydrodynamics of this scenario have been considered by \citep{Taam:1978dk,Bodenheimer:1984fs}. Accretion onto the NS during this phase is relatively efficient as neutrinos provide a cooling channel \citep{Houck:1991kc,Chevalier:1993by,Chevalier:1996kr,Fryer:1996kr}, leading to the suggestion that the NS might grow to collapse to a BH in some, but not all, cases \citep{Chevalier:1993by,Brown:1995jj,Bethe:1998jv,1998ApJ...502L...9F,Belczynski:2002gi,Kalogera:2007kh,Chevalier:2012gr}. 

\item {\it BH + 16 $M_\odot$ red supergiant (RSG):} In this scenario, a stellar mass BH interacts with its massive companion. This scenario might be realized under several circumstances. If, as suggested above, accretion-induced collapse from NS to BH ever occurs during CE evolution, that would leave a BH interacting with a massive-star envelope mid CE event. Perhaps more simply, if a massive star evolves and produces a stellar mass BH remnant, the post supernova orbit may lead to CE evolution \citep{Postnov:2014fb}. 

\end{itemize}

\begin{figure}[tb]
\begin{center}
\includegraphics[width=0.47\textwidth]{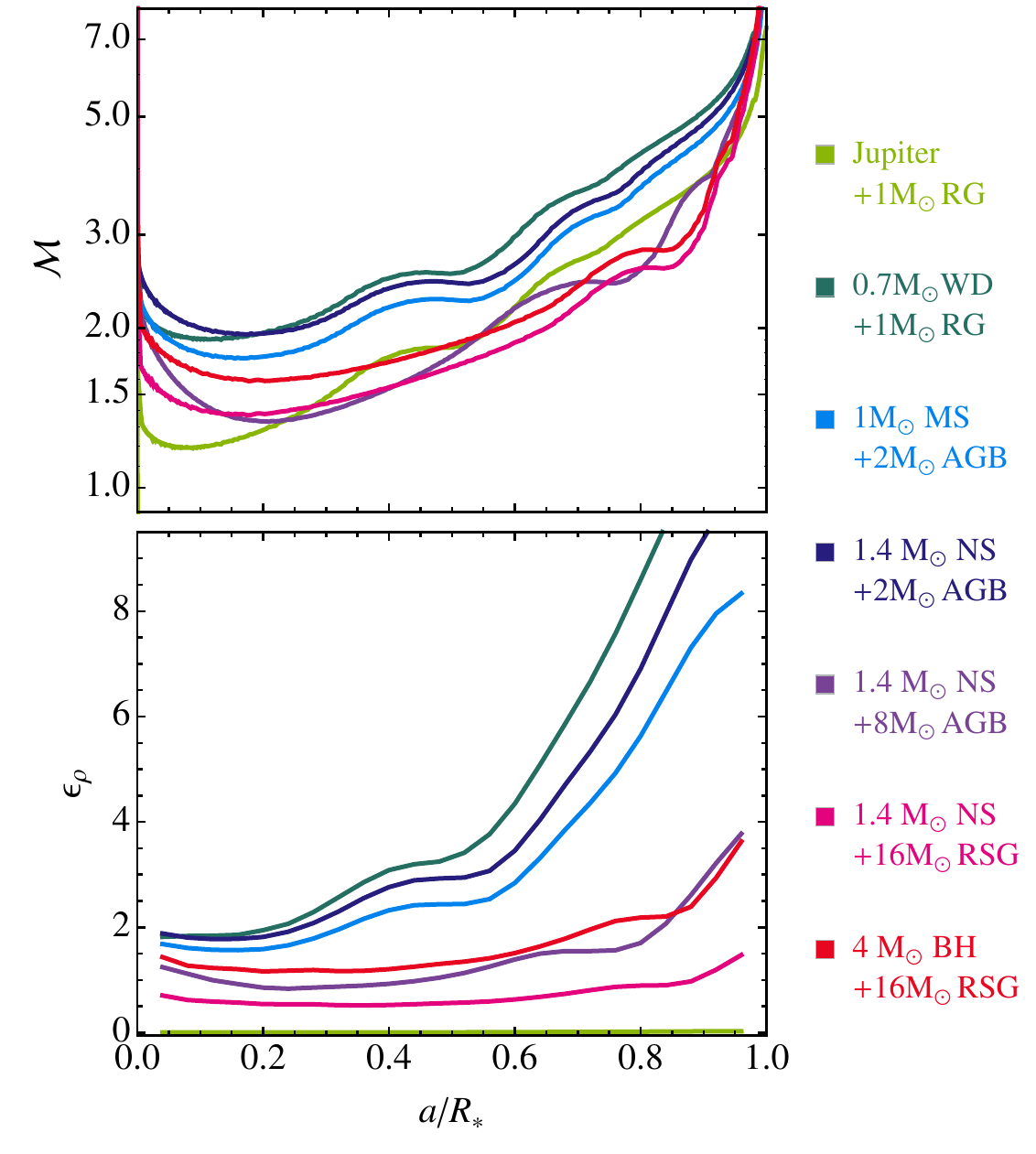}
\caption{Characteristic Mach numbers and density gradients for a variety of CE object pairings. Mach numbers are consistently highest and density gradients are steepest  near the limb of stars, where energy diffusion dictates that the density scale height becomes small. Typical density gradients span a wider range than  Mach numbers do, but for separations within the inner 50\% of $R_*$, $0.5 \lesssim \erho \lesssim 4$ are representative values for the embedded star and compact object cases.  }
\label{fig:scalesCompare}
\end{center}
\end{figure}

Figure \ref{fig:scalesCompare} is analogous to the upper and lower right-hand panels of Figure \ref{fig:scalesEvolution}, where we plot typical Mach numbers and gradients as a function of normalized radius. Rather than showing a time-evolution, we plot representative models for each encounter combination in 7 different representative CE pairings of objects. Mach numbers vary by at most a factor of a few at a given radius. This is not entirely unexpected because the highest Mach numbers are realized when the two objects are close to equal mass, while the lower limit occurs when the giant dominates the total mass. The gradients differ somewhat more significantly. In the extreme limit shown of a Jupiter-like planet embedded in a $1M_\odot$ giant, the gradient across the accretion radius is nearly zero. In general, gradients are steepest near the limb of the envelope because energy diffusion dictates that the density scale height becomes very steep at the photosphere. The most-nearly equal mass cases exhibit the steepest gradients, on average, in large part because the accretion radius tends to be a larger fraction of the stellar radius and thus encompasses a broader range of densities. 

\begin{figure*}[tbp]
\begin{center}
\includegraphics[width=0.45\textwidth]{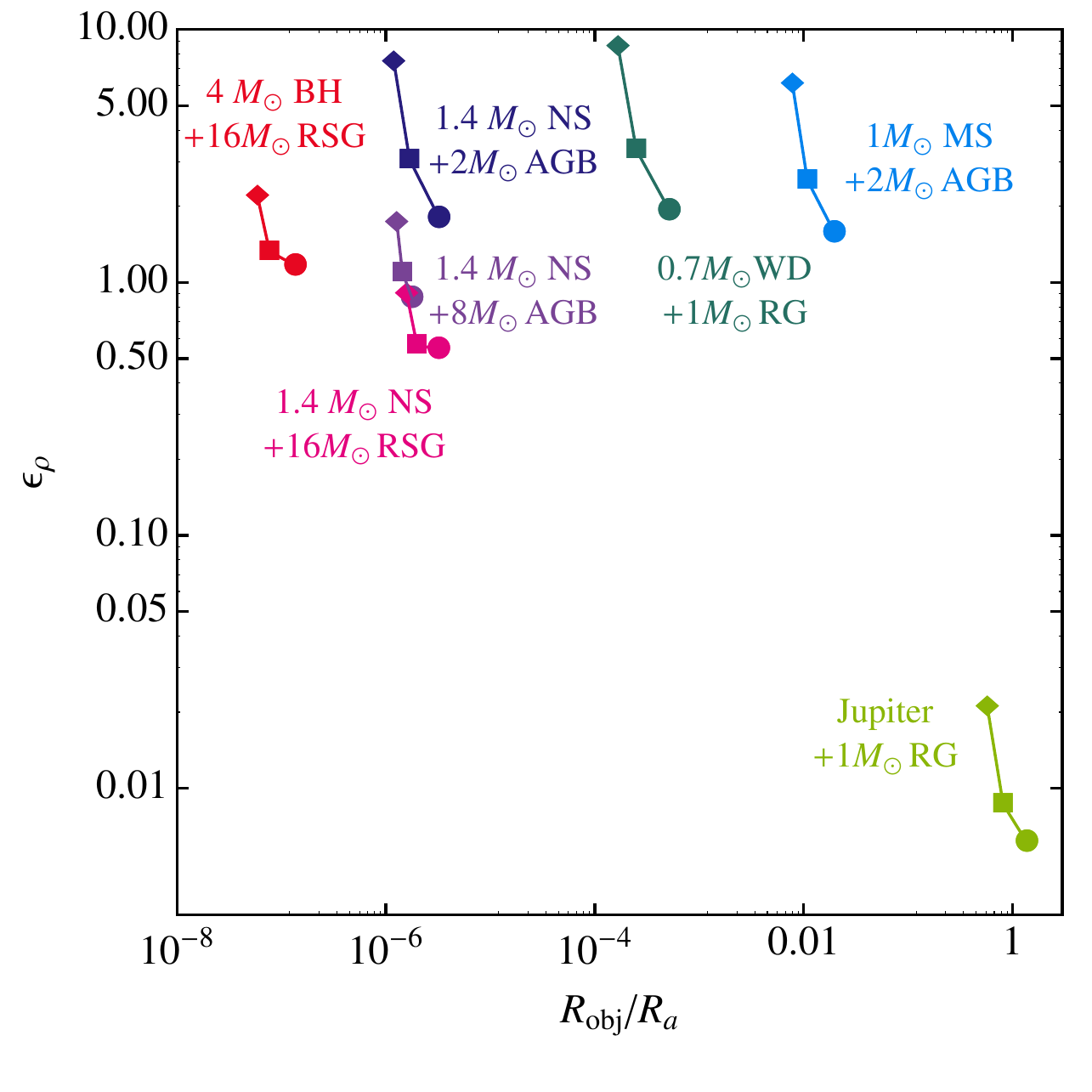}
\includegraphics[width=0.45\textwidth]{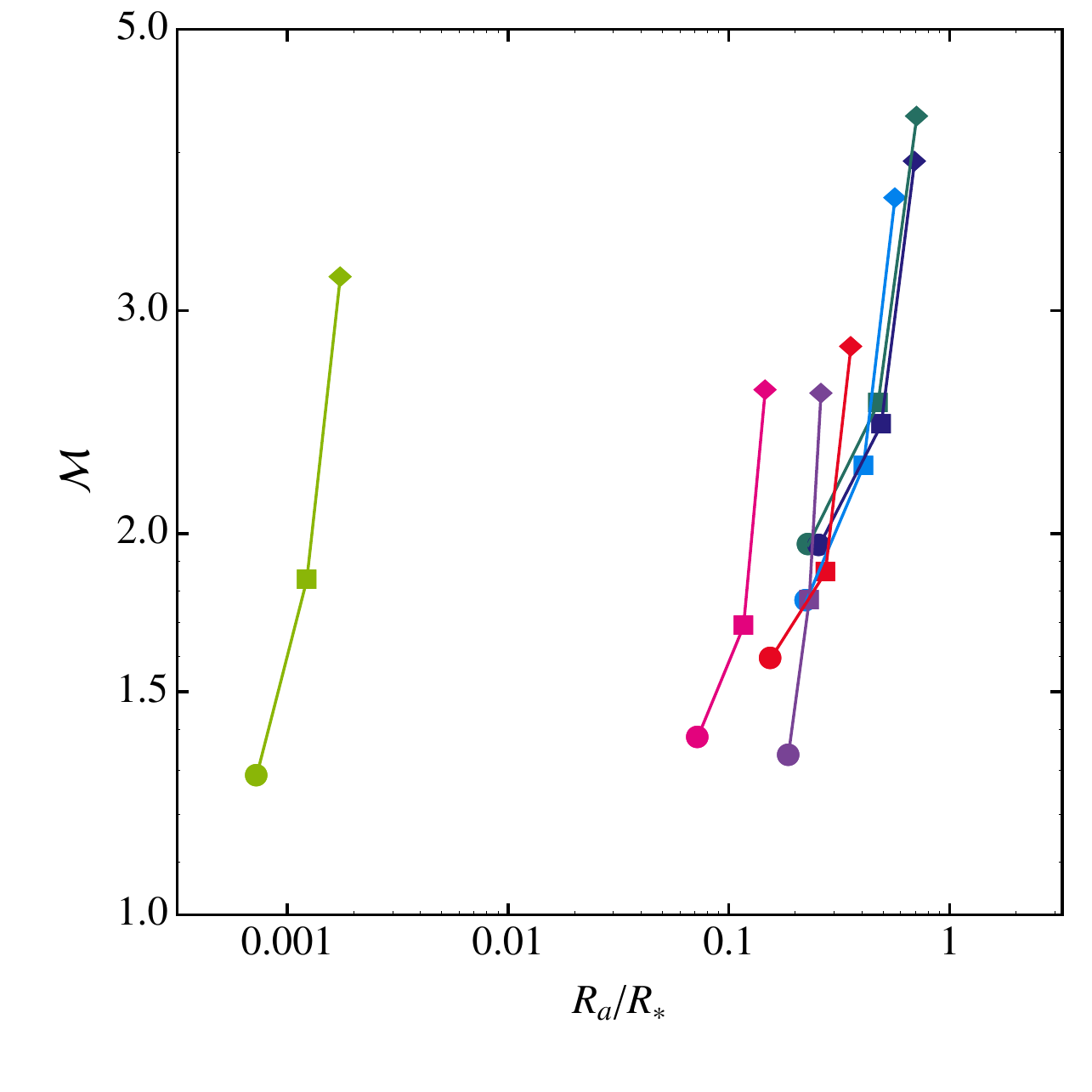}
\caption{Typical flow parameters plotted for the set of CE scenarios shown in Figure \ref{fig:scalesCompare}. Points evaluate the conditions in a given model at $0.2R_*$, $0.5R_*$ and $0.8R_*$ (with circles, squares, and diamonds, respectively). Colors are the same from upper to lower panels. The upper panel shows that embedded stars and compact objects are typically many times smaller than their gravitational capture radius. Only for an embedded Jupiter-like planet are the gravitational and geometric cross-sections similar.  The more massive objects are subject to significant density gradients, parameterized by $\erho$, while the planet is the only model with a mild gradient. The lower panel compares the local accretion radius to the stellar radius. While this ratio depends on the structural properties of the star, it is primarily determined by the mass ratio of the embedded object to its companion. Hydrostatic balance of the envelope structure ensures that typical Mach numbers are consistent, and mildly supersonic. These results are tabulated for $a=0.5R_*$  in Table \ref{scalestable}. }
\label{fig:scalesSummary}
\end{center}
\end{figure*}

The properties of these encounters are further summarized in Figure \ref{fig:scalesSummary}. In the upper panel we plot the density gradient as parameterized by $\erho$ against the radius of the embedded object compared to the accretion radius. The ratio $R_{\rm obj}/\Ra$ effectively compares the geometric ($\sim \pi R_{\rm obj}^2$) to the gravitational ($\sim \pi \Ra^2$) cross section. In the case of an embedded Jupiter-like planet, these scales are similar. However, for embedded stars and compact objects the accretion radius is typically $10^2$ to $10^7$ times larger than the object radius. The symbols plotted show the values evaluated at $a=0.2,0.5,$ and $0.8 R_*$. The lower panel of Figure \ref{fig:scalesSummary} shows the other two characteristic parameters, the Mach number, $\mach$, and the size of of the accretion radius as a fraction of the stellar radius, $\Ra/R_*$. 
The mass ratio of the embedded object to its companion is significant. In cases where the embedded object's mass is small compared to the envelope mass, the accretion radius is a small fraction of the total radius. In this case, fewer density scale heights are subtended by the accretion radius, leading to weaker density gradients, $\erho$. However, when the objects are relatively similar in mass, the accretion radius is of similar order of magnitude to the envelope radius, and it can therefore sweep across many density scale heights. These results are summarized in Table \ref{scalestable}, in which we give the numerical values evaluated at a single separation, $a=0.5R_*$. 

\begin{table*}[tbp]
\caption{Typical Encounter Properties from MESA Simulations  (evaluated at $0.5R_*$)}
\begin{center}
\begin{tabular}{lccccc}
\hline
\hline
Objects & Stellar Radius $R_* [R_\odot]$   & ${\mathcal M}$ & $\Ra/R_*$ & $R_{\rm obj}/\Ra$ & $\erho$ \\
 (1) & (2) & (3) & (4) & (5) & (6) \\    
\hline
Jupiter + $1M_\odot$ RG   & 103 &  1.84 &  $1.22(-3)$ & 0.82 & $8.64(-3)$ \\
$0.7 M_\odot$ WD + $1M_\odot$ RG   & 103 & 2.53 & 0.47 & $2.50(-4)$ & 3.34 \\
Sun + $2M_\odot$ AGB  & 226 & 2.26 & 0.41 & $1.09(-2)$ & 2.54 \\
$1.4 M_\odot$ NS + $2M_\odot$ AGB  & 226 & 2.44 & 0.49 & $1.69(-6)$  &  3.06 \\
$1.4 M_\odot$ NS + $8M_\odot$ AGB  & 562 & 1.77 &	0.23 & $1.44(-6)$  &  1.09 \\
$1.4 M_\odot$ NS + $16M_\odot$ RSG  & 807 & 1.69 & 	0.12 & $1.99(-6)$ & 0.56 \\
$4 M_\odot$ BH + $16M_\odot$ RSG  & 807 & 1.86 & 0.27 &	$7.69(-8)$  & 1.33 \\
\hline
\end{tabular}
\end{center}
\label{scalestable}
\end{table*}

\section{Methods}\label{sec:methods}
Our numerical approach is to perform idealized simulations based on the phase space of physically motivated flow parameters derived in the previous section. Our simulations follow closely in the tradition of simulations of HLA. We set up our experiment under many of the same premises to examine the effect of a substantial exponential density gradient in the upstream flow. In the following subsections, we first mention the most relevant previous numerical work on which our simulation builds, then discuss our numerical setup, and finally the simulation parameters chosen in our suite of runs.

\subsection{Previous Numerical Studies of HLA}

This work extends a long tradition of numerical study of HLA that began with the work of \citet{Hunt:1971vh}. 
The reader is directed to  \citet{Edgar:2004ip} a recent review, and to \citet{Foglizzo:2005df} for a detailed comparison of published simulations that attempts to synthesize  results with respect to flow stability as it depends on Mach number, accretor size, geometry, and equation of state.  Rather than restating their work here, we review only a few of the most relevant and recent studies. 

A benchmark series of 3D simulations with high spatial resolution and a variety of flow parameters was published beginning with \citet{Ruffert:1994je}. These simulations adopted homogeneous upstream boundary conditions. They employ nested grids to resolve a small region surrounding a point mass, from which a sphere is excised and a vacuum boundary condition is applied. Within this framework, flow morphologies, stability conditions, and accretion rates for different flow Mach numbers and equations of state were studied. \citet{Ruffert:1994cu} examined flows with adiabatic $\gamma=5/3$ and Mach number $\mach=3$.  They found that accretion rates in steady state were reasonably approximated by the HLA formula, but that after the flow developed in the first few crossing times, some loss of axisymmetry occurred. This imparted time variability to the accretion rate, but the dramatic flip flop instabilities seen in lower-dimensionality simulations were, strikingly, not observed in 3D \citep{Foglizzo:2005df}. 
Subsequent work \citep{Ruffert:1994vg,Ruffert:1995uf,Ruffert:1996wb} confirmed the more-stable configurations of 3D flows, and found that more violently unstable scenarios occurred with smaller accretion boundaries, higher Mach numbers, and more compressible equations of state. 

These simulations were recently revisited in 2D \citep{Blondin:2009ji,Blondin:2013hm} and 3D \citep{Blondin:2012ep,Naiman:2011ip,Toropina:2011ci,Lee:2014im}. Again, major differences in flow morphology are found between 2D and 3D geometry. In particular, rotationally supported flows develop in 2D planar geometries while these are not observed in 3D. \citet{Blondin:2012ep} study two different accretor sizes with $\gamma=5/3$ and $\mach=3$ to find extremely stable accretion with the larger accretor (with radius 5\% of the accretion radius), but a breathing mode instability and some variation in accretion rate with an accretor of only 1\% of $\Ra$. Very little loss of axisymmetry is observed in these simulations, however, with differences as compared to \citet{Ruffert:1994cu} potentially attributable to differences in grid meshing and resolution. 

Of particular importance to this study is work that has extended the traditional HLA problem to look at background gradients of velocity and density.  \citet{Livio:1986tr}, \citet{Soker:1986tg}, \citet{Livio:1986vr} studied accretion from an inhomogeneous medium using a 3D particle in cell hydrodynamics method. Although the resolution around the accretor with this method is limited, they found that the introduction of gradients lead to the accretion of only a small amount of the angular momentum available in the upstream flow. \citet{Fryxell:1988kb} and \citet{Taam:1989da} studied density and velocity gradients, respectively, in 2D planar accretion with a grid-based hydrodynamics method. This approach vastly improved the ability to study the flow just outside the accretor. \citet{Fryxell:1988kb} adopted $\gamma =4/3$, and found that small gradients of density lead to a flip-flop instability \citep{Livio:1991uo} of rotation around the accretor as vortices are shed into the wake. With steeper gradients, \citet{Fryxell:1988kb} displaced structures in the wake and very limited accretion rates. 
\citet{Armitage:2000eg} carry out relatively similar simulations except with a hard surface rather than inflow central boundary condition and evaluate the degree of rotational support of material bound to the accretor.  \citet{Ruffert:1997wy} and \citet{Ruffert:1999tq} applied a mild upstream gradient which varied by 3\% or 20\% in either velocity or density was applied to their 3D accretion setup of \citet{Ruffert:1994je}. In these simulations, the inhomogeneous upstream conditions lead to unsteady rotation of the flow  but not to the formation of steady disks surrounding the accretor with $\gamma=5/3$.   Some accretion of angular momentum from the post shock region resulted, but a dramatic modification in the average accretion rate was not observed.

\begin{table*}[t]
\caption{FLASH simulation parameters and results }
\begin{center}
\begin{tabular}{lcccccccc}
\hline
\hline
Name &  ${\mathcal M}$ & $\erho$ & $\gamma$ & $\Rs/\Ra$ & $\Rs/\delta_{\rm min}$ &  $\dot M/\mhl$ & $F_{{\rm d},x}/(\pi \Ra^2 \rhoinf \vinf^2)$  \\ 
\hline
A & 2  & 0 &  $5/3$ & 0.05 & 12.8 & 7.05(-1)& 1.98  \\
B & 2  & 0.1&$5/3$ &  0.05 & 12.8 & 6.03(-1) & 1.89  \\
C & 2  & 0.175 &$5/3$ &  0.05 & 12.8 & 5.04(-1) & 1.74 \\
D & 2  & 0.3 &$5/3$ &  0.05 & 12.8  & 2.66(-1) & 1.57\\
E & 2  &0.55 &  $5/3$ &0.05 & 12.8  & 1.27(-1) & 1.48 \\
F & 2  & 1 &  $5/3$ &0.05 & 12.8  & 4.09(-2) & 1.18 \\
G & 2  & 1.75 &  $5/3$ &0.05 & 12.8  & 2.56(-2) & 1.58 \\
H & 2  & 3 &  $5/3$ &0.05 & 12.8  & 2.74(-2) & 3.05 \\
\hline
I & 2  & 0 & $5/3$ &  0.01 & 10.2  & 6.41(-1) & 1.94\\
J & 2  & 0.1 & $5/3$ &  0.01 & 10.2  & 3.78(-1) & 1.71 \\
K & 2  & 0.175 & $5/3$ &  0.01 & 10.2  & 2.51(-1) & 1.58 \\
L & 2  & 0.3 & $5/3$ & 0.01 & 10.2  & 1.53(-1) & 1.44 \\
M & 2  & 0.55 & $5/3$ &  0.01 & 10.2  & 9.79(-2) & 1.45 \\
N & 2  & 1 &  $5/3$ &0.01 & 10.2  & 2.02(-2) & 1.19  \\
O & 2  & 1.75 & $5/3$ &  0.01 & 10.2  & 1.44(-2) & 1.58 \\
P & 2  & 3 &  $5/3$ &0.01 & 10.2  & 1.09(-2) & 3.93 \\
\hline
Q & 2 & 0.3 & $5/3$ & 0.05 & 6.4 & 2.64(-1) & 1.56 \\
R & 2 & 0.3 & $5/3$ & 0.05 & 25.6 & 2.69(-1) & 1.57\\
S & 2 & 0.3 & $5/3$ & 0.05 & 51.2 & 2.69(-1) & 1.57 \\
\hline
T & 1.1 & 0.3 & $5/3$ &0.05 & 12.8 & 1.90(-1) & 4.71(-1) \\ 
U & 1.1 & 1 & $5/3$ &0.05 & 12.8 & 5.19(-2) & 8.98(-1) \\
\hline
V & 3 & 5 & $5/3$ &0.01 & 10.2 & 6.37(-3) & 17.6 \\
X & 3 & 5 & $1.1$ &0.01 & 10.2 & 1.03(-1) & 18.0 \\
\hline
\end{tabular}
\end{center}
\label{simstable}
\end{table*}

\subsection{Numerical Approach and Simulation Setup}

While previous work has examined the effects of mild upstream gradients in velocity and density on the HLA problem, we have shown in Section \ref{sec:scales} that in many cases the relevant density gradients may be substantially stronger than those studied by  \citet{Ruffert:1999tq}. 
Thus we extend this work and perform 3D hydrodynamic simulations of accretion flows using the FLASH code \citep{Fryxell:2000em}.  We solve the fluid equations using FLASH's directionally split Piecewise Parabolic Method Riemann solver \citep{Colella:1984cg}. We make use of a gamma-law equation of state, and in most cases use $\gamma =5/3$.  
A 3D cartesian grid is initialized surrounding a point mass that is fixed at the coordinate origin. 

These simulations are performed in dimensionless units, where $\Ra = \vinf = \rhoinf = 1$.
In these units, the characteristic time is $\Ra/\vinf=1$, and the characteristic accretion rate $\mhl = \pi$, equation \eqref{eq:mhl}.  
 The $-x$ boundary feeds a wind of material into the box and past the point mass. 
A simulation is then parameterized by the upstream Mach number, $\mach$, measured at $y=z=0$. 
We also allow for a planar density gradient in the $\hat y$ direction, parameterized by $\erho = \Ra / \Hrho$. The $\pm y$ and $\pm z$ boundaries are outflow boundaries positioned at $\pm 10\Ra$ (10 in code units). On the downstream, $+x$, boundary (positioned at $+4\Ra$) we apply a diode boundary condition that allows for outflow but not inflow. 

We create a spherical absorbing boundary condition ``sink'' surrounding the central point mass, with radius $\Rs$. 
The potential of the point mass is smoothed, but only well within this sink, at a radius of $\approx\Rs/10$. Thus, this smoothing does not affect the flow outside the excised region.
 Each time step, the average pressure and density are computed within a shell that extends from $\Rs$ to $2\Rs$. Then, the pressure and density inside the sink are set to $10^{-3}$ of these surrounding values to create a vacuum that does not impinge on the surrounding flow. This vacuum condition represents accretion with no feedback on the surrounding flow. Material in the vicinity of the sink is allowed to expand into it at the sound speed and thus be absorbed \citep{Ruffert:1994je}. 
 
 Before the pressure and density are overwritten within the sink, we integrate the accreted mass and angular momentum. The mass accretion rate is then defined as the accreted mass divided by the time step, $\dot M = \delta m /{\text dt}$; the components of the angular momentum accretion rate are defined similarly.  As a consistency check, we performed tests with sink density and pressure pre-factors between $10^{-1}$ and $10^{-5}$  with no visible difference in the accretion of mass or angular momentum.  However, a qualitatively different central boundary condition, particularly one that did apply a feedback on the flow, would likely result in different mass and angular momentum accumulation rates. We explore the effects of such a boundary condition in Section \ref{sec:feedback}.  

The initial conditions are of constant velocity in the $\hat x$ direction, where the $x$-velocity is $v_x = \vinf$. The initial density field, $\rho_i$, is a function of the $y$ position as
\beq\label{rhofunc}
\rho_i = \rhoinf \exp(\erho y),
\eeq
applied only within $-2<y<2$ to limit the total range of density in the background material of the computational volume. Within this planar gradient of densities, high densities are found at $+y$ coordinates. We turn the point mass on progressively over the first time unit, so that it is fully active after $\Ra/\vinf=1$.

The simulations employ the PARAMESH library to provide adaptive mesh refinement to resolve small features around the accretor within the large simulation box \citep{2000CoPhC.126..330M}.  Adaptive refinement is based on the second derivative of pressure. The box is initialized with 7 blocks (of $8^3$ cells) in the $x$ direction, and 10 each in the $y$ and $z$ directions.  We then allow for between 6-9 levels of adaptive refinement of those coarsest blocks. To avoid devoting all of the computational effort to features far from the accretor, we force the maximum level of refinement allowed for a given block to drop in proportion to the radius from the coordinate origin (where the accretor resides). Blocks with size less than $\alpha r$, where we adopt $\alpha=0.3$, are not allowed to be further refined. The first decrement in refinement occurs at $r\approx3 \Rs$, and drops one level further each time the radius doubles \citep[See][for another astrophysical problem in which this refinement criteria is applied in FLASH]{2014ApJ...785..123C}.

\subsection{Simulation Parameters}

The primary effect we explore in this paper is the inclusion  of a significant density gradient to the upstream flow in supersonic accretion. To that end we perform  a series of simulations with increasing density gradient $\erho$. We adopt a Mach number of $\mach=2$ for these simulations, and a gas ratio of specific heats, $\gamma=5/3$, that is representative of an ideal gas flow in which cooling is ineffective -- like that embedded deep within a stellar interior. In one series, simulations A--H, we adopt a sink boundary condition size of $\Rs=0.05\Ra$. In simulations I--P, we reduce the sink size to $\Rs=0.01\Ra$.  
These $\Rs=0.05\Ra$ simulations are run from $t=0$ to $t=80 \Ra/\vinf$. The $\Rs=0.01\Ra$ are started from checkpoints of the $\Rs=0.05\Ra$ simulations with the same upstream conditions at $t=20\Ra/\vinf$ and run for $10\Ra/\vinf$. 
We perform a resolution study, simulations Q--S, which all have $\mach =2$, $\Rs=0.05\Ra$ to demonstrate the robustness of our derived accretion parameters. 
We perform some simulations with lower and higher Mach number, $\mach=1.1$ (T,U) and $\mach=3$ (V,X) to test the sensitivity of flow morphology to upstream Mach number. In simulation X, we adopt $\gamma=1.1$ to represent a flow which is can cool more effectively than the adiabatic conditions.

\section{Results}\label{sec:results}

The introduction of an upstream density gradient breaks the symmetry that defines classical HLA. In the following subsections we explore the effects of this symmetry-breaking on the morphology, accretion rates, sink-size effects, rotation, and drag realized our hydrodynamic simulations. 

\subsection{Flow Morphology}

\begin{figure*}[p]
\begin{center}
\includegraphics[width=\textwidth]{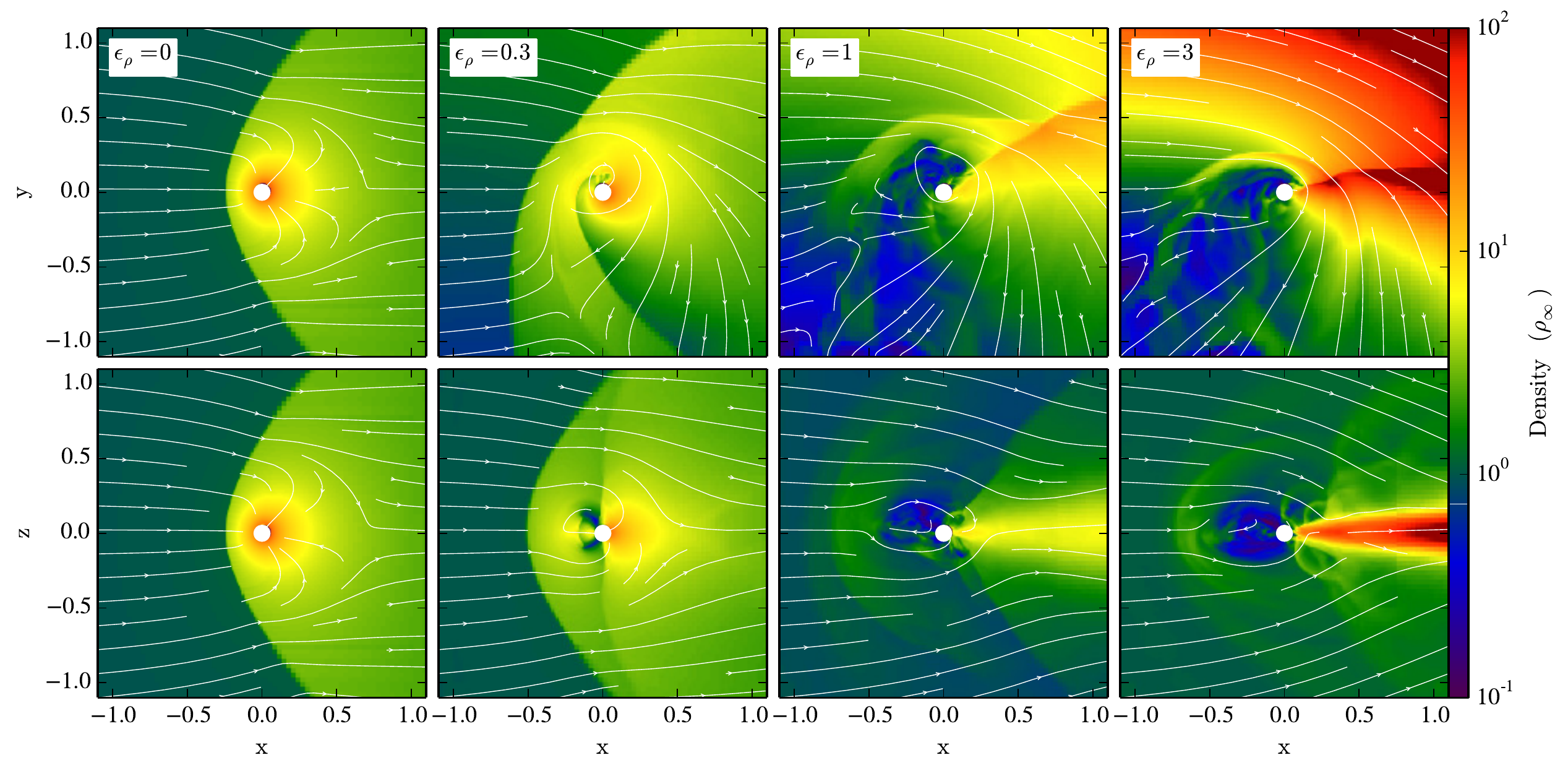}
\vspace{0.5cm}
\includegraphics[width=\textwidth]{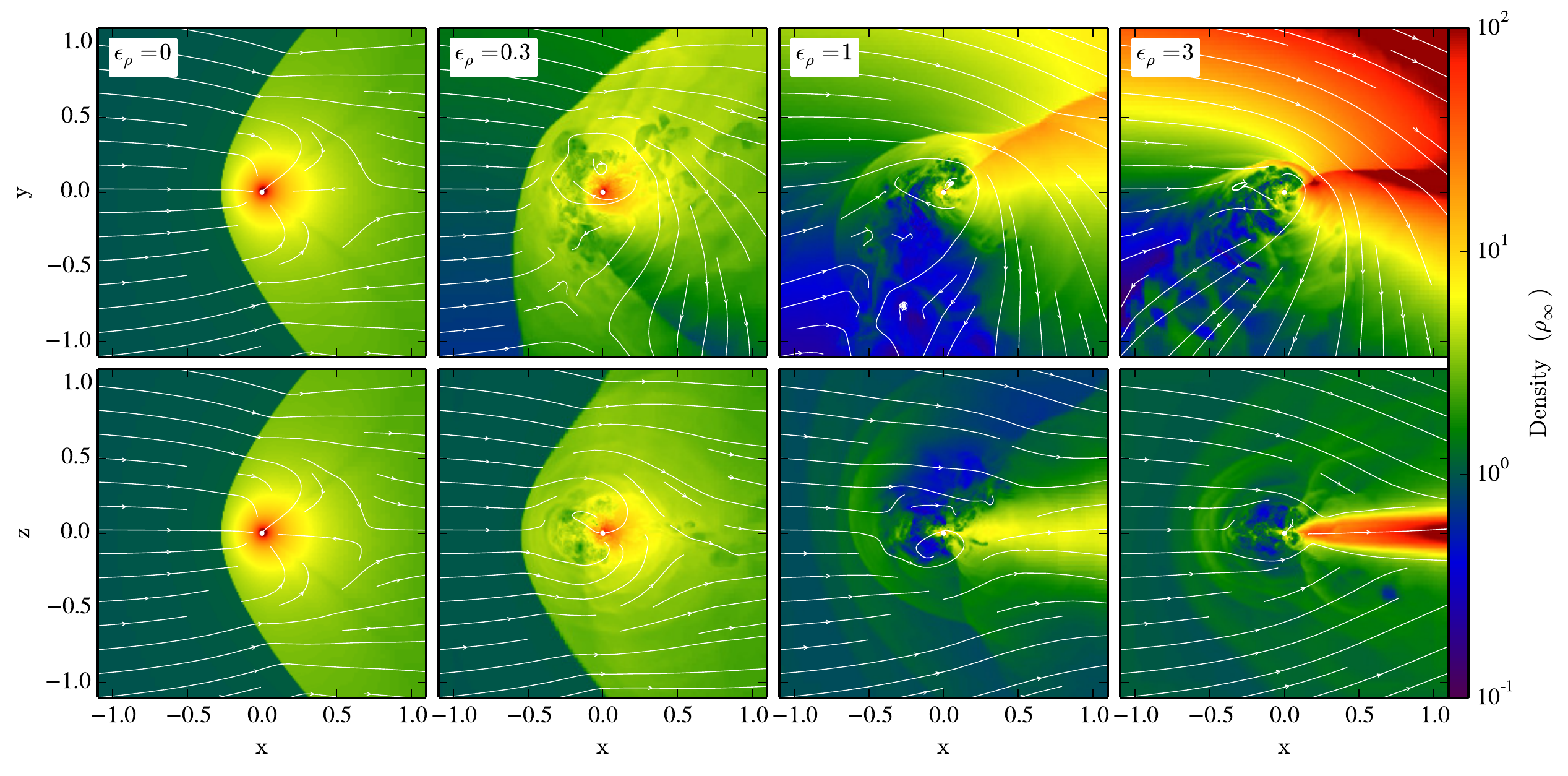}
\caption{Comparison of flow morphologies realized with different upstream density gradients. The upper series of panels shows simulations with $\Rs=0.05\Ra$ (simulations A, D, F, and H). The lower series has $\Rs=0.01\Ra$ (simulations I, L, N, and P). Density is plotted in terms of $\rho_\infty$, the density at $y=0$, while axis labels are in units of $\Ra$. The upstream Mach number is $\mach=2$ in all simulations, and snapshots are shown at a time  $30 \Ra/\vinf$.  As gradients are introduced, increasing asymmetry develops in the imposed plane of rotation ($x$-$y$). Bow shock structures migrate from symmetric to tilted to finally wrapping nearly completely around the accretor into the wake for $\erho=3$.  Dense material (at positive $y$ coordinates) is focused around the accretor and wraps past the accretor in the wake to impinge on lower-density material. In the perpendicular plane ($x$-$z$), dense material is increasingly concentrated in the wake as the gradient steepens. Dense features in the wake are complimented by low-density pockets of turbulent material upstream. This and following figures visualizing FLASH simulation fields are made using the {\texttt yt} toolkit \citep{Turk:2010dd}. 
}
\label{fig:sim_grad_dens}
\end{center}
\end{figure*}

The introduction of upstream density gradients introduce dramatic changes to the morphology of  the flow around objects embedded within a CE. Figure \ref{fig:sim_grad_dens} displays these changing flow morphologies for upstream gradients of $\erho = 0, 0.3,1,3$ and sink sizes of $\Rs = 0.05\Ra$ and $\Rs=0.01\Ra$. We show two slices through the simulation domain, one in the $x-y$ plane, the same plane as the imposed $\hat y$ density gradient, and one in the $x-z$ plane, perpendicular to the imposed gradient. 

In the zero gradient, HLA-case, the $x-y$ and $x-z$ slices are nearly identical. A high degree of symmetry is preserved in this case of homogeneous upstream conditions. Flow lines converge toward a stagnation region in the wake of the accretor and some material reverses to fall into the sink from this accretion column \citep[e.g.][]{Edgar:2004ip}. Most of the accretion, therefore, occurs in the downstream hemisphere of the accretor \citep{Blondin:2012ep}. 

The symmetry of the $\erho=0$ case is broken by the introduction of an upstream gradient. Although the sink itself is small with respect to the density scale height, $\Rs\ll \Hrho$, the bow shock sweeps through a large density contrast, affecting the flow even at small scales near the accretor. With 0.3 density scale heights per accretion radius, ($\erho = 0.3$, the second panel in Figure \ref{fig:sim_grad_dens}) the flow morphology is distorted and it presents a tilted bow shock structure to the upstream flow. As the density gradient steepens further, the bow shock continues to rotate to face the flux of densest material. As a result, the bow-shock is nearly reverse-facing by the time the gradient steepens to $\erho=3$, the right-hand panel of Figure \ref{fig:sim_grad_dens}. Further, the single shock interface of the symmetric case is replaced by multiple nested shocks  at different rotation angles with respect to the accretor. In the $\erho=1,3$ cases with $\Rs=0.05$, a one-sided trailing shock facing the high density material extends inward to the surface of the accretor. When the sink size is reduced to $\Rs=0.01\Ra$, a low density cavity forms surrounding the accretor and the tail shock does not remain attached. 

In these steep gradient cases, material of high and low density are both focused toward the wake of the accretor from positive and negative $y$ coordinates, respectively. The momenta of these fluid parcels do not cancel, however, as occurs in the case of homologous upstream conditions. Thus, the density gradient also introduces a net angular momentum swept up by the bow shock, and material in the post shock region carries net rotation around the accretor. This effect is observed in the flow lines overplotted on Figure \ref{fig:sim_grad_dens}. A downstream accretion column cannot exist in this scenario, because flow in the wake region is moving  tangentially with respect to the accretor. 

\begin{figure*}[p]
\begin{center}
\includegraphics[width=\textwidth]{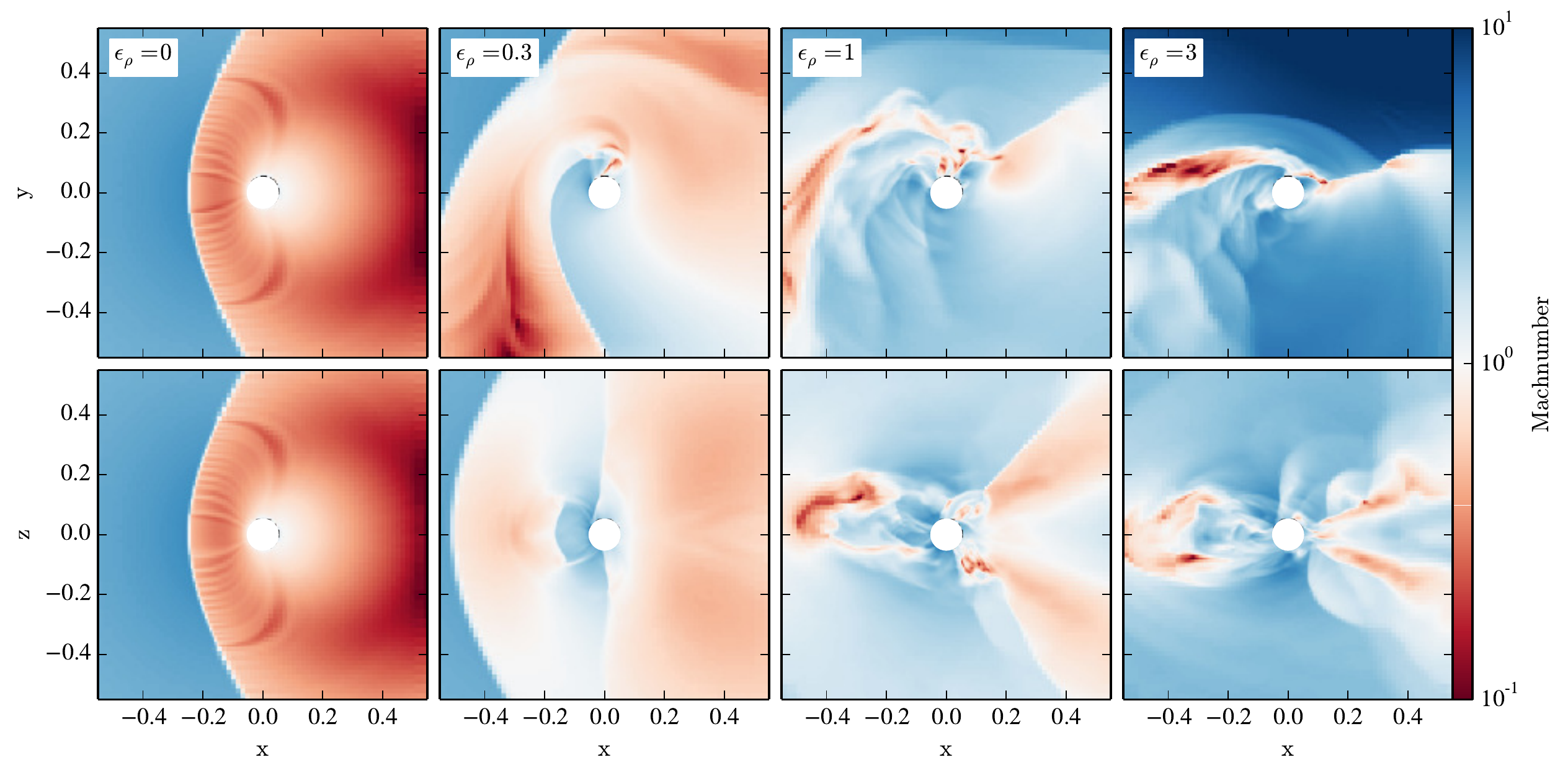}
\vspace{0.5cm}
\includegraphics[width=\textwidth]{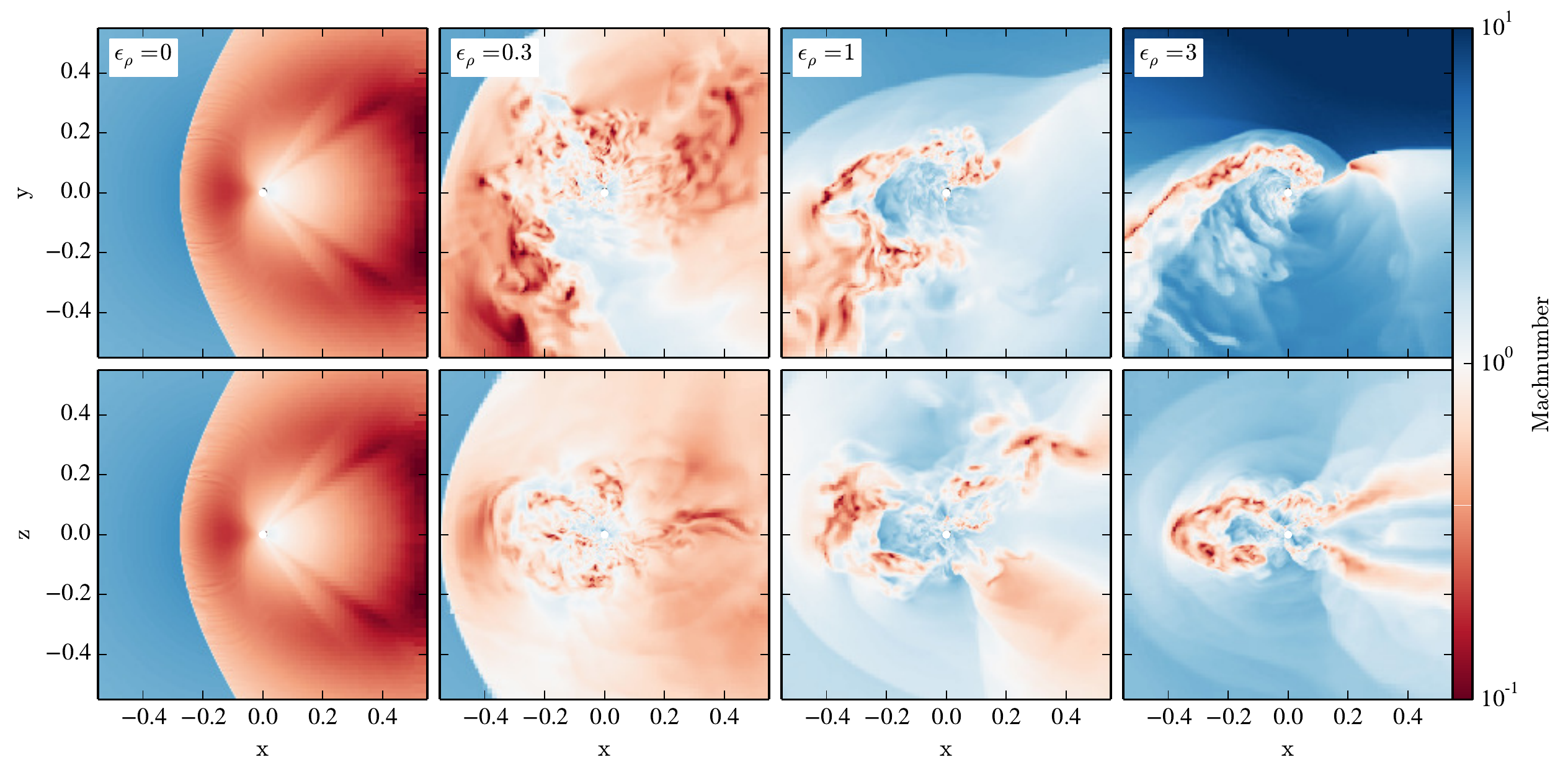}
\caption{Flow Mach number for the same frames shown in Figure \ref{fig:sim_grad_dens}  (simulations A, D, F, and H). Steepening gradients (and smaller sink sizes) lead to increasingly turbulent flows. Boundary layers are of shock heated, and therefore low-$\mach$, material divide regions of flow moving in different directions. Angular momentum is redistributed in these shocks with some plumes of material falling inward extending inward to the accretor, while other material is swept outside of $\Ra$. In the zero-gradient case, some spurious features representative of the cartesian discretization arise. These artificial features are swept away by continuous fluid motion as soon as any gradient is introduced.   Note the axis scale is a factor of two closer here than in Figure \ref{fig:sim_grad_dens}.   }
\label{fig:sim_grad_mach}
\end{center}
\end{figure*}

In the absence of velocity cancelation, flow lines show that much of the dense material is never focused into the accretor. This result was anticipated by \citet{Dodd:1952uc} who, in analytically calculating the capture cross section of the accretor given a linear upstream density gradient, note that captured material need not go on to accrete. Instead, they state that only the material whose angular momentum can be redistributed could be expected to fall into an accretor.  Traces of this process are seen in Figure \ref{fig:sim_grad_mach}. In particular, only material whose tangential velocity is partially canceled in the nested shock structures can fall into the sink.

Figure \ref{fig:sim_grad_mach} examines the flow Mach numbers for the same set of simulation snapshots as Figure \ref{fig:sim_grad_dens}. An extended subsonic region trails in the wake in the symmetric HLA case, while a sonic surface inside the bow shock marks the reacceleration of stalled material toward the accretor \citep{Foglizzo:1997vq,Foglizzo:1999wp,Foglizzo:2005df,Blondin:2012ep}. In our $\erho=0$ panel of Figure \ref{fig:sim_grad_mach}, some numerical artifacts can be seen originating from grid interfaces at the bow shock. These flaws highlight the difficulty of simulating a quasi-steady flow with a cartesian grid mesh. With symmetry broken as the upstream gradient is introduced, these artifacts disappear as the flow is no longer so near to steady state. 

With the introduction of a density gradient and the corresponding lack of cancelation of tangential motion, material in the post-bow shock region carries some rotational support. This rotation leads to the development of regions surrounding the accretor that retain supersonic velocities. In contrast to the $\erho=0$ simulation, where the leading face of the accretor remains in sonic contact with the bow shock, for $\erho>0$, the sonic surface largely detaches from the accretor. The cavity of subsonic material in which sound waves can propagate becomes increasingly restricted in the steeper gradient simulations. The flow asymmetry and instability observed in the  steep density gradient cases cannot therefore be due to acoustic perturbations cycling from the bow shock to the accretor, as is thought to drive instability  for sufficiently high Mach number flows with homologous upstream conditions and $\gamma =5/3$ \citep{Foglizzo:2005df}.  Instead, the observed  instability  appears to be a vortical instability seeded in  flow near the embedded object  that carries too much angular momentum to accrete, and instead impinges on the surrounding material. This is related  to the  instability observed in isothermal flows with nearly-homologous upstream boundary conditions, which also tend to develop significant rotation \citep{Foglizzo:2005df}. Shear layers and density inversions with respect to the accretor's gravity can be seen to develop within these flow structures and these effects also seed instability and vorticity in the post bow shock region. In general, as the gradient steepens and the sink size decreases, more unstable and turbulent flow is exhibited surrounding the accretor. 

It is worthwhile here to examine some of the ways that these flows may be compared to previous numerical work with inhomogenous upstream density. Flow that breaks axisymmetry and develops significant rotation is  observed in simulations with both velocity and density gradients by \citet{Ruffert:1997wy} and \citet{Ruffert:1999tq}. Our simulation in the second panel of Figure  \ref{fig:sim_grad_dens} ($\erho=0.3$) may be compared morphologically with model "NS" of \citet{Ruffert:1999tq}, which is the steepest-gradient model explored and has $\Ra/\Hrho=0.2$ (see their Table 1 for parameters, and Figure 2 for a plot of the density distribution and flow vectors). These models exhibit very similar morphology, and in particular, similar flow patterns in the post shock region. As a point of comparison, it is worth noting that in our setup higher densities are at positive $y$ values, while they are at negative $y$ values in \citet{Ruffert:1999tq}. By contrast, 2D cylindrical simulations with upstream density gradients by \citet{Fryxell:1988kb} and \citet{Armitage:2000eg} show qualitatively different behavior. Even with mild density gradients these 2D flows develop small, rotationally supported disks \citep{Armitage:2000eg} trailed by an attached wake that is unstable in the transverse  sense and sheds vortices from the accretion region \citep{Fryxell:1988kb}. We explore this difference further in Section \ref{sec:disks}.

\subsection{Effects of Sink Size}\label{sec:sinksize}

\begin{figure}[tbp]
\begin{center}
\includegraphics[width=0.45\textwidth]{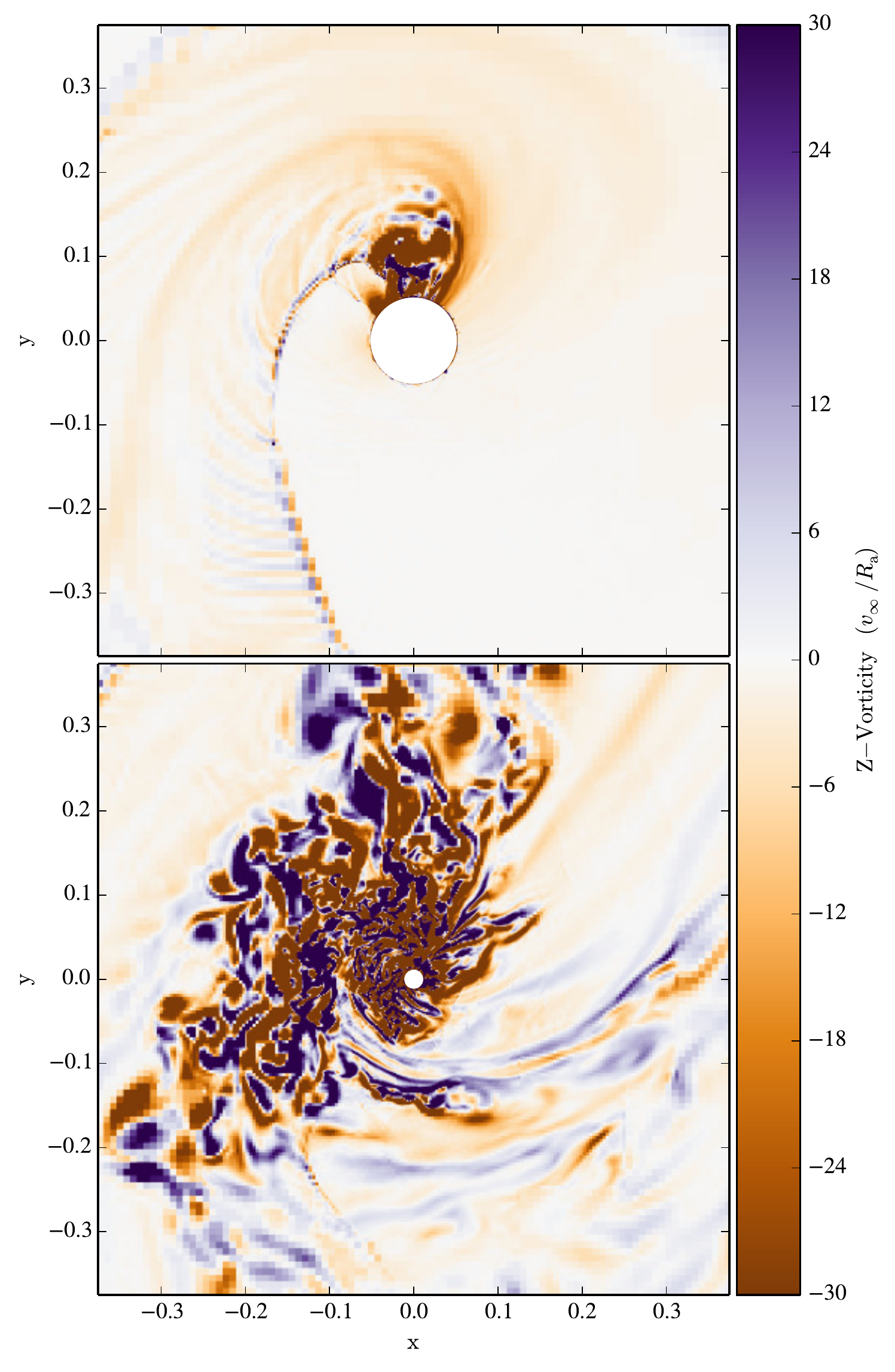}
\caption{Comparison of the $z$-component of the vorticity in the inner regions of simulations for which $\Rs=0.05\Ra$ and $\Rs=0.01\Ra$  (simulations S, top, and L, bottom). Both have otherwise identical parameters, with a gradient of $\erho=0.3$, and identical maximum linear resolution $\delta_{\rm min} \approx 10^{-3} \Ra$. Imposing the sink boundary in the flow at smaller radii allows material to penetrate deeper into the accretor's gravitational potential without being subsumed. While the flow for $\Rs=0.05\Ra$ is mostly laminar, strong local vorticity develops surrounding the accretor when $\Rs=0.01\Ra$. The extent of turbulent flow extends well beyond the accretor size in either simulation and becomes an important feature of the large-scale flow.  }
\label{fig:vort}
\end{center}
\end{figure}

A comparison of the upper and lower panels of Figures \ref{fig:sim_grad_dens} and \ref{fig:sim_grad_mach} highlights some differences in the large-scale flow that result from the size of the accretor. This effect is explored further in Figure \ref{fig:vort} for the $\erho=0.3$ case of simulations D and L. While rotating flow is relatively laminar in the $\Rs=0.05\Ra$ simulation, vortices dominate the region in the $\Rs=0.01\Ra$. 
At the root of this difference is that, with the introduction of a density gradient in the upstream flow, there is an angular momentum barrier as well as an energetic barrier to accretion.

We can estimate the radius at which material with impact parameter $<\Ra$ will circularize if it completely inelastically redistributes its momentum. Under these assumptions, the specific angular momentum (in the $\hat z$ direction) specified by the density distribution at infinity is $l_{z,\infty} = \dot L_z(<\Ra) / \dot M (<\Ra)$, where 
\beq\label{dotMRa}
\dot M(<\Ra) = \vinf \int_{<\Ra} \rho(y) dA.
\eeq
Here the area element can be re-written $dA=2\sqrt{\Ra^2-y^2}dy$ with limits $y=\pm \Ra$. This expression reduces to $\mhl$ when $\erho=0$. Similarly,
\beq\label{dotLRa}
\dot L_z (<\Ra) =  2 \vinf^2 \int_{-\Ra}^{\Ra}  y  \rho(y)  \sqrt{\Ra^2-y^2}   dy.
\eeq
Given equation \eqref{rhofunc} for $\rho(y)$, we can solve analytically for $l_{z,\infty}$ and find 
\beq
l_{z,\infty} = \frac{I_2(\Ra \erho)}{I_1(\Ra \erho)} \Ra \vinf
\eeq
where $I_n$ are the modified Bessel functions of the first kind,
\beq
I_n(z) = {1 \over \pi} \int_0^\pi e^{z \cos \theta} \cos (n \theta) d \theta, 
\eeq
for integer $n$ \citep{1972hmfw.book.....A}. The radius at which material with this angular momentum will be rotationally supported is $R_{\rm circ} = l_{z,\infty}^2 / G M$. In our adopted units, $\Ra =1$ and $\vinf=1$, $GM=1/2$, we thus have
\beq\label{rcirc}
R_{\rm circ} = 2 \left[\frac{I_2(\erho)}{I_1(\erho)}\right]^2.
\eeq
This quantity represents the radius at which the fluid with impact parameter less than the accretion radius will circularize if it has the opportunity to perfectly cancel momenta with opposing flow. 

If $R_{\rm circ} < \Rs$, material can enter the accretor directly, sweeping through at most one orbit.  Visually, this is what we observe in the $\Rs=0.05\Ra$ case of Figure \ref{fig:vort}. Material that is drained from the accretion region acts as an effective cooling source on the flow in that accretion removes material from the vicinity on of the accretor so that it does not proceed to impinge on newly-inflowing material. Instability arises and vorticity appears to grow in our simulations when $R_{\rm circ} > \Rs$. In this case, there is an angular momentum barrier to accretion, and material is trapped in orbit around the accretor. This scenario becomes inherently unstable when coupled to the continuous infall of fresh material from the upstream region. The new material collides with the old as it tries to penetrate to the accretor. Shearing layers and density inversions that result near the accretor lead to instabilities that amplify such that turbulence dominates the post-shock region. 
Numerically, equation \eqref{rcirc} suggests a circularization radius of $R_{\rm circ}(\erho=0.3) = 0.011$ for the example in Figure \ref{fig:vort}.  This is mildly outside $\Rs=0.01$ and substantially inside $\Rs=0.05$ indicating the significance of the transition observed in Figure \ref{fig:vort}. We should, therefore, not expect convergence of the flow behavior with respect to sink size in flows where angular momentum plays a role in shaping flow morphology. This is particularly true in cases with $R_{\rm circ}\sim \Rs$.

\subsection{Accretion of Mass and Angular Momentum}

In our simulations we track the accumulation of mass and angular momentum into the central sink boundary. Each time step the accumulated quantities above the floor state are integrated over the sink volume before the pressure and density are re-written. The rate implied is calculated each time step by $\dot X = \Delta X / dt$, where $X$ is an arbitrary quantity, $\Delta X$ is the integrated new material, and $dt$ is the time step.

\begin{figure}[tbp]
\begin{center}
\includegraphics[width=0.45\textwidth]{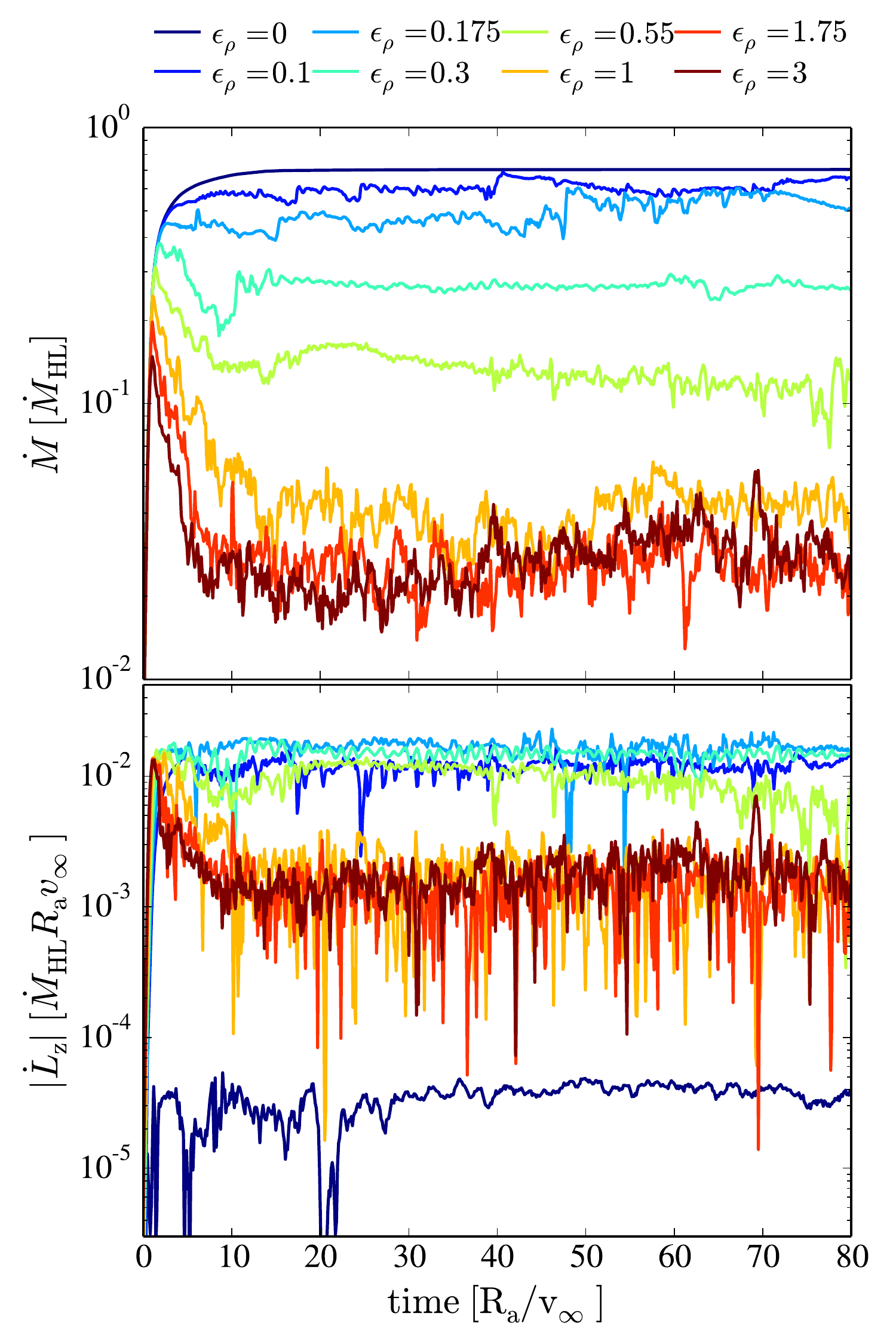}
\caption{Accretion of mass (top) and angular momentum (bottom) as a function of time in  simulations with a sink size $\Rs =0.05\Ra$  (simulations A-H).  Accretion rates are normalized to $\mhl$ (for mass, equation [\ref{eq:mhl}]) and to $\mhl \Ra \vinf$ (for angular momentum).  The upper panel shows that the introduction of an upstream gradient not only dramatically decreases the accreted mass but also leads to increased  chaotic time-variability compared to the median value. The variability observed can be attributed to turbulence in the flow seeded by the upstream gradient. The accretion rate of angular momentum is lowest in the $\erho=0$ simulation, and provides a measure of the high degree to which our simulations preserve symmetry intrinsic to the setup. The $\erho=0.1-0.55$ cases are able to accrete substantially more angular momentum than the stronger-gradient cases because $\dot M$ is not so highly impeded. Rotation is always found in the sense imposed by the upstream gradients.   }
\label{fig:sim_acc}
\end{center}
\end{figure}

In Figure \ref{fig:sim_acc}, we show the time-dependent accretion of mass (upper panel) and angular momentum (lower panel) for the series of simulations (A--H), for which $\Rs=0.05\Ra$. We run these simulations for $80\Ra/\vinf$, but accretion rates relax to their steady-states within the first domain-crossing time $\approx10\Ra/\vinf$ as found, for example by \citet{Ruffert:1994je}. As has been demonstrated in previous numerical simulations, in the zero-gradient case $\mhl$ provides a good order-of-magnitude estimate of the accretion rate \citep[e.g.][]{Ruffert:1994je,Ruffert:1994cu,Naiman:2011ip,Blondin:2012ep}.
As the upstream density gradient steepens, the steady state accretion rate drops precipitously. Interestingly, the early time accretion rate in all cases is similar to that of the zero gradient case. The accretion rate tracks that of the homogeneous case and breaks off only when the bow shock has swept wide enough to trace out substantial upstream inhomogeneity and angular momentum. 
In cases where transient flows exist in which the bow shock is less than fully developed, the effective density gradient is thus reduced in proportion to the bow shock's extent. 
As the upstream density gradient steepens, the accretion rate also becomes increasingly variable. The variability seen is chaotic and there is no single apparent periodicity or driving timescale in a Fourier decomposition of $\dot M(t)$ \citep[e.g.][]{Edgar:2005bi}. 

The lower panel of Figure \ref{fig:sim_acc} compares the accretion rate of angular momentum between simulations with differing density gradients. Here we normalize our results  to a characteristic angular momentum accretion rate, $\mhl \Ra \vinf$. The zero-gradient case preserves symmetry to better than 1 part in $10^4$, and provides a gauge for the fidelity of the other cases. 
 As the gradient appears ($\erho<1$), the accreted angular momentum at first increases because of rotation imparted to the post bow shock flow. For steeper gradients, the accreted angular momentum actually decreases again because $\dot M$ decreases in the steepest-gradient cases. For these combinations, the limiting of $\dot M$ appears to outweigh the increasing angular momentum content of the upstream flow. 

\begin{figure}[tbp]
\begin{center}
\includegraphics[width=0.45\textwidth]{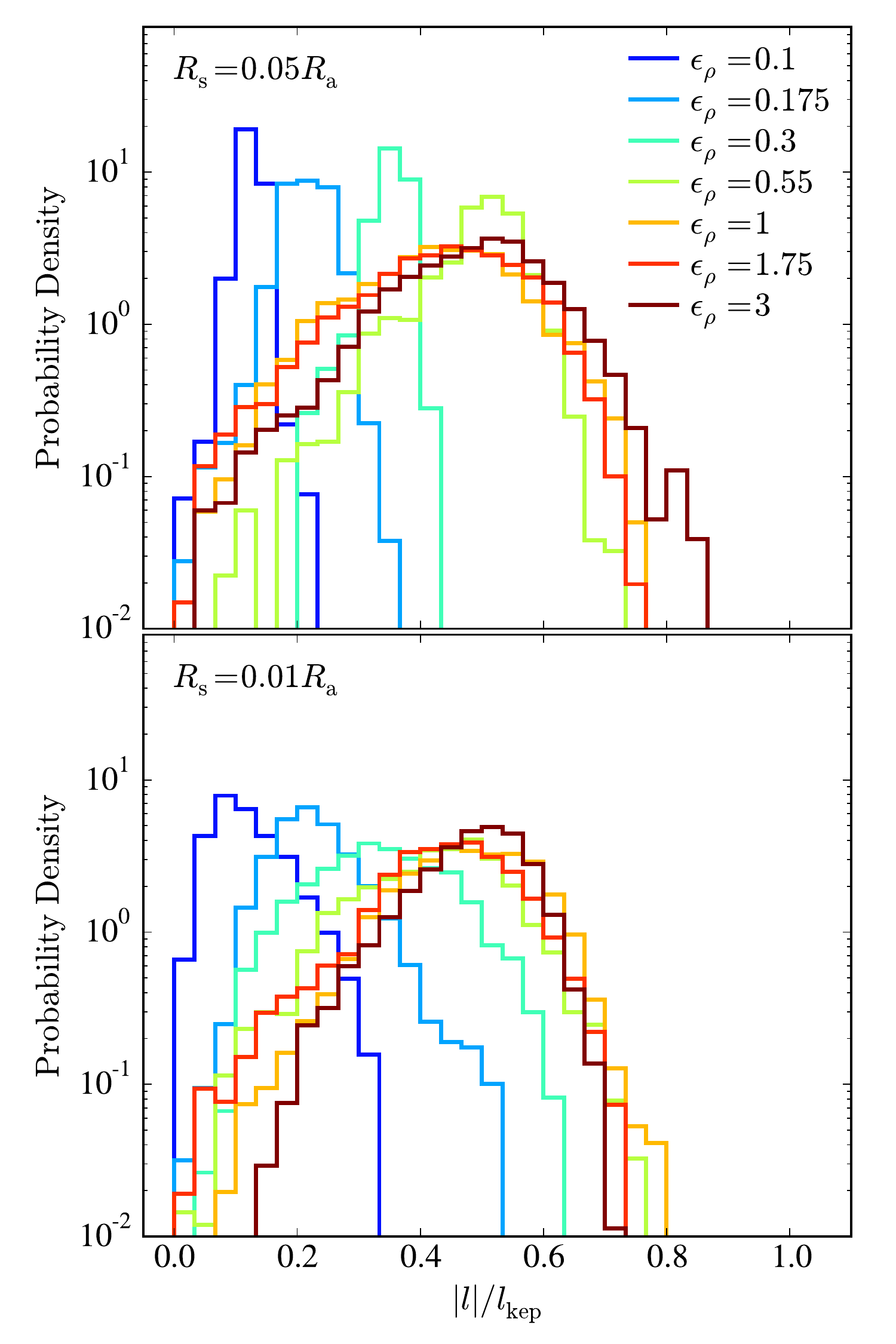}
\caption{Specific angular momentum of accreted material as compared to the angular momentum of material in Keplerian rotation at the sink radius in simulations A-H (top) I-P (bottom). Here $l_{\rm kep} = \sqrt{G M \Rs}$, in the dimensionless units of the simulations, $GM=1/2$ and this becomes $l_{\rm kep} = \sqrt{\Rs/2}$.  
A transition can be seen in both panels as the density gradient steepens. The angular momentum content of accreted material at firs increases, but then maximizes at a mean value of $|l|/l_{\rm kep} \approx 0.5$.
In the steeper-gradient cases ($\erho >0.55$) the width of the distribution also broadens. However, accreted material always displays substantially sub-Keplerian rotation. 
The transition noted in Figure \ref{fig:vort} can be seen above by comparing the $\erho=0.3$ case in the upper and lower panels. In the upper panel, where $\Rs=0.05\Ra$, the accreted angular momentum forms a narrow distribution, similar to that exhibited in the milder-gradient cases. In the lower panel, in which $\Rs=0.01\Ra$, the distribution is broader, nearly joining the family of curves from the steep-gradient cases. As mentioned in the caption of Figure \ref{fig:vort}, this transition takes place material circularizes inside (for $\Rs=0.05\Ra$) or outside (for $\Rs=0.01\Ra$) the sink radius. 
}
\label{fig:specificangularmomentum}
\end{center}
\end{figure}

\begin{figure}[tbp]
\begin{center}
\includegraphics[width=0.45\textwidth]{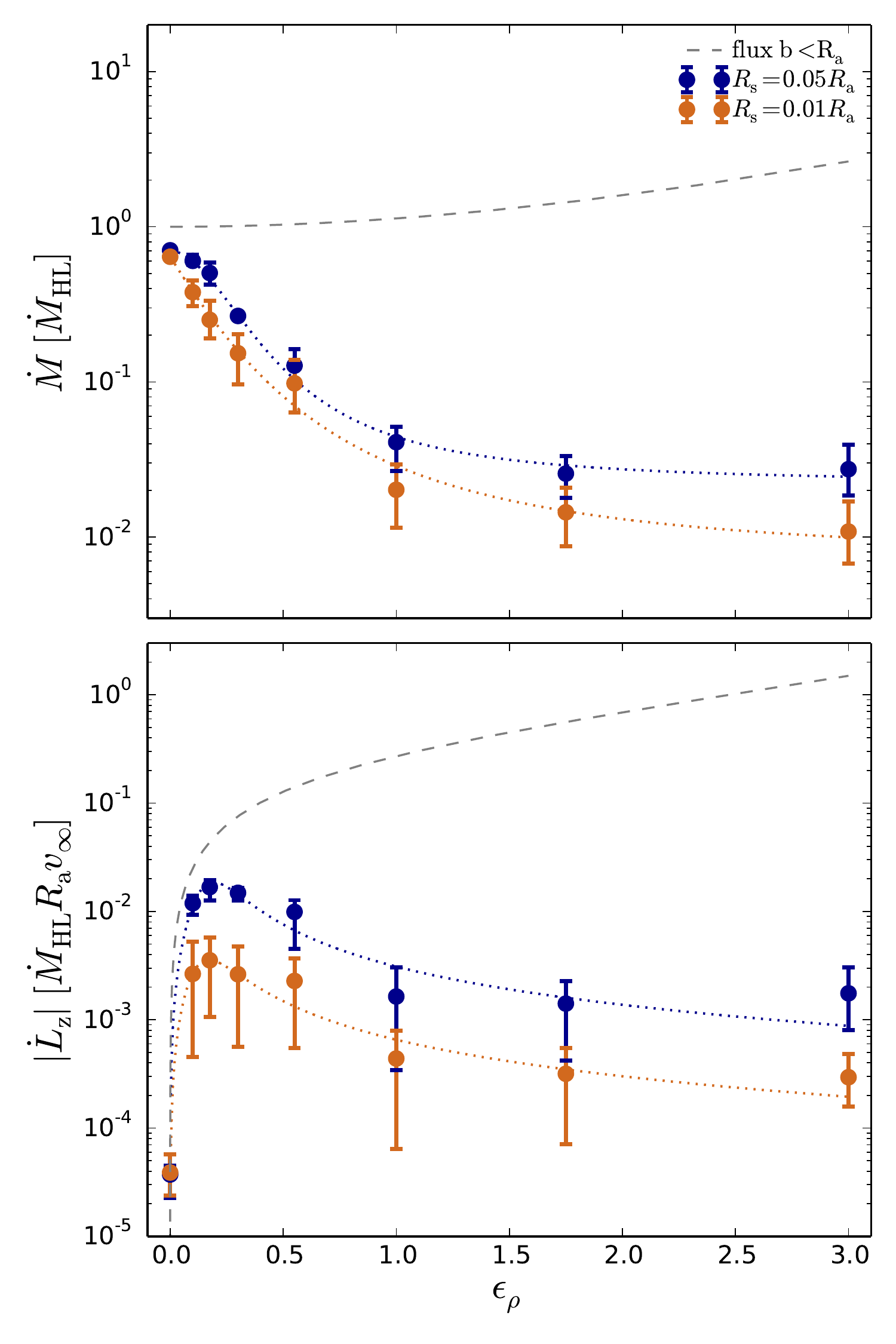}
\caption{Accretion of mass and angular momentum summary with respect to gradient, $\erho$ in simulations A-H and I-P. The points show the median, while the error bars show the 5\% and 95\%   percentile bounds of the  time series data for different simulations, calculated for times $t>20\Ra/\vinf$ when the simulations are in steady-state. The dashed lines show the flux of either mass or angular momentum through a surface with impact parameter at infinity, $b <\Ra$. Although the available mass and angular momentum increase as an exponential density gradient is introduced,  accretion is dramatically inhibited by the asymmetric flow geometry that develops. The smaller-sink simulations exhibit lower mass and angular momentum accretion rates with larger variability.   }
\label{fig:accsummary}
\end{center}
\end{figure}

To disentangle the accretion rate of mass and angular momentum, we turn our attention to the specific angular momentum of accreted material. In Figure \ref{fig:specificangularmomentum}, we plot histograms of the specific angular momentum content of accreted material.  The specific angular momentum, $l = \sqrt{\dot L_x^2 + \dot  L_y^2 +  \dot L_z^2} / \dot M$,   is normalized to the angular momentum of material in Keplerian orbit at the sink radius, $l_{\rm kep} = \sqrt{G M \Rs}$. With $\Ra=\vinf=1$ as used in the simulations, $GM=1/2$ and this becomes $l_{\rm kep} = \sqrt{\Rs/2}$. Even in the gradient cases, we find that the angular momentum accretion rate is much less than Keplerian. 
The specific angular momentum content of accreted material is also highly variable, as visualized by the broad histograms in Figure \ref{fig:specificangularmomentum}.  Interestingly, $|l|$ is similar in $\erho=0.55-3$ simulations, despite the increased flux of angular momentum from the upstream conditions. The milder gradient simulations peak at somewhat lower typical angular momentum content $|l|/l_{\rm kep} \approx 0.1-0.4$ than the steeper-gradient simulations which peak at $|l|/l_{\rm kep} \approx0.5$.
This transition in behavior occurs when the circularization radius is outside the sink, $R_{\rm circ} \gtrsim \Rs$, as opposed to when the net angular momentum allows circularization inside the sink. 
 The lack of a tail extending to $|l|/l_{\rm kep} \approx1$ indicates that none of the simulations ever reach a state of accretion from a nearly-Keplerian flow. It is worth contrasting this, briefly, to recent 2D planar simulations in which, after an initial growth phase, the specific angular momentum of accreted material is nearly always within a few percent of the Keplerian value \citep{Blondin:2009ji,Blondin:2013hm}. This difference in accretion modality, therefore, appears to lie in the geometry of the simulations.

Figure \ref{fig:accsummary} summarizes the accretion of mass and angular momentum in our simulations. 
We have plotted the accreted mass and angular momentum for simulations with $\Rs=0.05\Ra$ and $\Rs=0.01\Ra$. 
We plot the flux of either mass or angular momentum through an upstream cross section with impact parameter $b<\Ra$, equations \eqref{dotMRa} and \eqref{dotLRa}. The mass and angular momentum available in the flow both increase with gradient. 
With $\erho =0.3$, the flux of mass with $b<\Ra$ is $1.01\mhl$, with $\erho =1$ it is   $1.13 \mhl$, while with $\erho=3$ it increases to  $2.64 \mhl$. 
The mass that reaches the sink and accretes decreases dramatically as the density gradient steepens. This limiting of accretion, despite the increased availability of material,  must be attributed to the change in flow structure and angular momentum barrier to accretion described in the previous subsections. 

The comparison between  $\Rs=0.05\Ra$ and $\Rs=0.01\Ra$ series results depends strongly on the upstream gradient imposed. The mass accretion rates decrease by a factor of a few with the small sink as compared to the large for the strongest-gradient cases, yet only by $\sim10$\% for zero-gradient. This difference points explicitly to the role of angular momentum in limiting the amount of material that is able to actually accrete. Figure \ref{fig:specificangularmomentum} supports this conclusion; the angular momentum content of accreted material nicely follows the normalization with respect to $l_{\rm kep}$, which depends explicitly on the sink radius.

\subsection{Disk Formation?}\label{sec:disks}

\begin{figure*}[tbp]
\begin{center}
\includegraphics[width=0.95\textwidth]{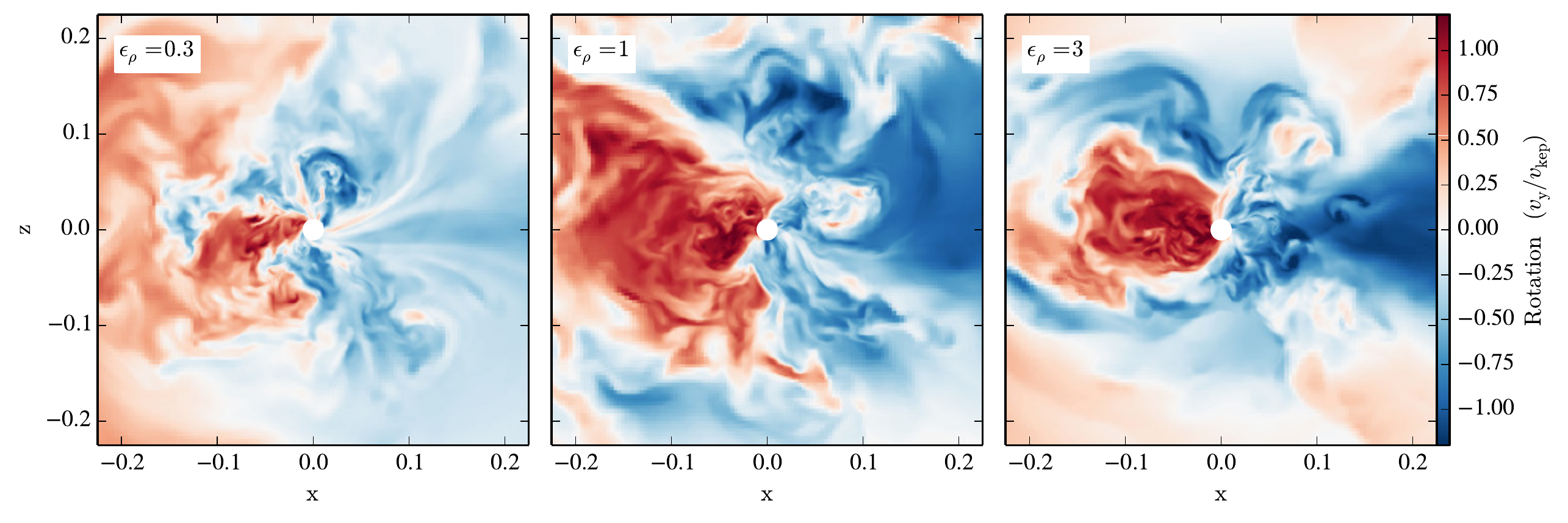}
\includegraphics[width=0.95\textwidth]{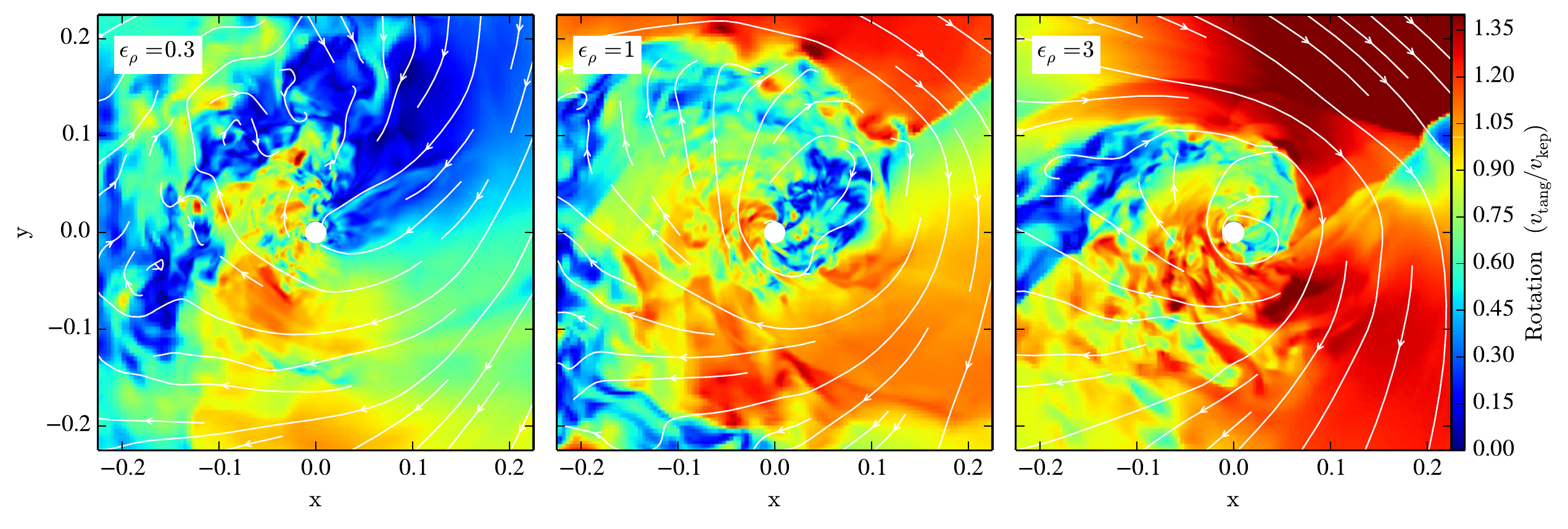}
\caption{Rotation imposed by an upstream density gradient  in simulations L, N, and P. In the upper panel, shown perpendicular to the plane of rotation, color shows inward (red) or outward (blue) motions compared to the local Keplerian velocity. The extent of the rotating material becomes smaller as the gradient increases and inflowing high-density material impinges on the orbiting material.  The lower panels look at the same frames and the magnitude of tangential motion in the orbital plane. These show that the rotational flow that develops is not strictly disk-like. Rotating material in the orbital plane is interspersed with cavities of material with pressure, rather than rotational, support. Plumes of material with little rotational support extend to the sink surface, feeding accretion.}
\label{fig:disks}
\end{center}
\end{figure*}

With the accretion of angular momentum apparent in simulations with an upstream gradient, a natural question that arises is whether the rotation imposed in the flow leads to the formation of persistent accretion disks. Disk formation is, for example, apparent in accretion flows that develop from wind-capture in widely spaced binaries \citep[see e.g.][for numerical simulations]{Zarinelli:1995wn,Blondin:2013hm,HuarteEspinosa:2013kw}. 
When disk structures form, viscosity from magnetorotational instability \citep{Balbus:1991fi} can transfer angular momentum to funnel disk material toward the central body. 
In these cases, the flow is primarily optically thin, and can cool effectively through line emission and the blackbody continuum. The effective polytropic index of the equation of state is assumed to be close to one, $\gamma \sim 1$, and as a result the flow is nearly isothermal. With effective cooling, rotationally supported structures are quickly assembled. 
When cooling is not effective, as is the case when embedded deep within a shared stellar envelope, the adiabatic index of the gas remains close to $\gamma =5/3$. In this case, the adiabatic build-up of pressure is significant as gas moves to smaller radii. Without a means to dissipate this internal energy, the flow remains in large part pressure supported.

Figure \ref{fig:disks} explores the rotational motion of material close to the accretor in simulations with an imposed gradient. The lower panels of this Figure look at tangential motion in the orbital plane, while the upper panel plots motion into and out of the perpendicular plane. In the $\erho=0.3$ simulation, rotational motion is sub-Keplerian nearly everywhere. In the larger gradient simulations there are pockets of material rotating with the Keplerian velocity. However, these pockets are interspersed with shocked material with little rotational support. Rather than exhibiting steady circular flow patterns, streamlines are highly elliptical in the orbital plane, being first flung upstream then wrapping around to encounter incoming material. These flow patterns define a cavity of material bounded by shocks in which angular momentum is redistributed. 
As visualized by flow streamlines, some material encountering these boundary layers is deviated back toward the accretor, but much of the rest is advected away and shed in the wake. That persistent orbits do not exist indicates that material does not have the opportunity to be viscously accreted before it is advected away from the sink region. 

The lack of a dense built-up disk feature is contrary to what has been observed in 2D cylindrical coordinates with a $\gamma=5/3$ equation of state. \citet{Fryxell:1988kb} and \citet{Armitage:2000eg} find rotationally-supported flows with the introduction of upstream density gradients in their 2D simulations. \citet{Blondin:2009ji} and \citet{Blondin:2013hm} look specifically at disk formation in 2D simulations of HLA. They find that even without any imposed gradients the flow is unstable to the development of quasi-Keplerian disks. As mentioned earlier, accreted material in these simulations typically carries specific angular momentum close to the Keplerian value ($|l|/l_{\rm kep} \approx 1$), which may be contrasted to Figure \ref{fig:specificangularmomentum}, in which we find $|l|/l_{\rm kep} \approx 0.5$ to be much more representative. In both of these sets of 2D simulations spiral shocks appear to mediate the transport of angular momentum that allows material in the disk to accrete \citep{Blondin:2000ez,Blondin:2013hm}.

Pressure certainly plays a role in distinguishing 2D and 3D simulations. Because of the difference in radial dependence of the volume element, material is compressed to differing degrees in 2D and in 3D. The standoff shock seen in 2D simulations is entirely  rotationally developed, being a consequence of the flip-flop instability saturating and wrapping around the accretor into the upstream flow \citep{Blondin:2009ji}. By contrast, a bow shock forms promptly in 3D simulations with $\gamma=5/3$ solely due to compression of convergent flow. This offers some explanation of why 2D adiabatic and 3D isothermal simulations show similar properties. The 2D volume element $dV_{\rm 2D}/dr=2\pi r$, while in 3D, $dV_{\rm 3D}/dr=4\pi r^2$. If we examine the adiabatic increase in pressure of accreting gas in three cases in which simulations have been performed,  $P \propto \rho^\gamma \propto V^{-\gamma}$. With $\gamma =1$ in 3D, $P\propto r^{-3}$, and disk formation is apparent. In 2D with $\gamma=5/3$ disks are again apparent, and $P\propto r^{-10/3}$. While in 3D with $\gamma=5/3$ purely-Keplerian disks do not appear, and $P\propto r^{-5}$.

\begin{figure}[tbp]
\begin{center}
\includegraphics[width=0.49\textwidth]{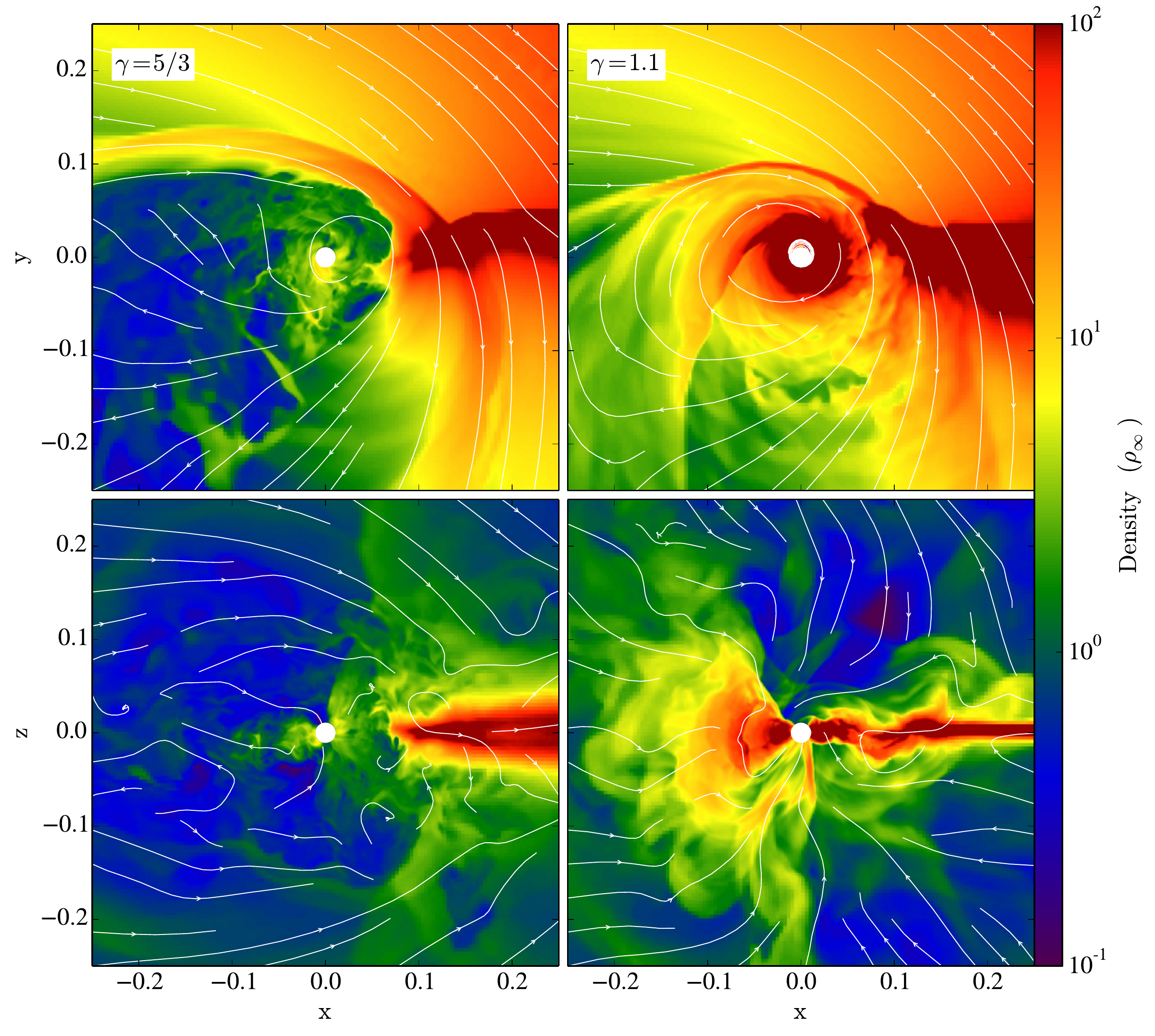}
\caption{A comparison between $\mach =3$, $\erho=5$ simulations with different equations of state. These simulations (V and X in Table 2), have $\gamma=5/3$ and $\gamma=1.1$, respectively. Just as in the other $\gamma=5/3$ simulations disk formation is not apparent in simulation V, despite the steep density gradient. In simulation X ($\gamma=1.1$), the more compressible equation of state permits the formation of thin, dense, and persistent disk.  }
\label{fig:steepeos}
\end{center}
\end{figure}

We illustrate this point with a comparison between simulations V and X in Figure \ref{fig:steepeos}, which share $\mach = 3$, $\erho = 5$, and $\Rs = 0.01 \Ra$ but differ in compressibility of the equation of state. Simulation V has $\gamma = 5/3$ while simulation X has $\gamma=1.1$.  While the $\gamma=5/3$ case forms the same cavity seen in the other steep gradient simulations, the $\gamma =1.1$ case forms a persistent thin disk. This disk is much denser than the surrounding material, and it sets up in steady rotation around the accretor. Notably, the scale height of the disk is thicker in the upstream direction as it is impinged on by incoming material. The $\gamma=1.1$ equation of state indicates that some cooling occurs in the gas, and thus is not physically realized deep within a CE. Early in the CE phase, though, near the surface layers of a star, the cooling time may be short and $\gamma<5/3$ is possible.

\subsection{ Drag }

The rate of momentum dissipation due to gravitational  focussing of  the surrounding gas sets the drag force felt by the embedded body. In turn, this corresponds to the rate of orbital energy dissipation into the CE and the rate that the embedded object inspirals to   tighter orbital separations. 

We estimate the  momentum dissipation realized in our simulations along the direction of orbital motion ($\hat x$) as follows, 
\beq\label{drageq}
F_{{\rm d},x} = \int \rho v_{x} (v_x \hat x \cdot  \hat r) dS
\eeq
where $F_{{\rm d},x}$ is the $x$-component of the drag vector, $\hat r$ is the vector normal to the surface of a unit sphere, and the surface of integration $dS$ is a sphere  with the gravitational radius of the accretor, $\Ra$ \citep[after][equation 3]{Ricker:2007kc}. 
We calculate this flux using the marching cubes surface reconstruction algorithm in \texttt{yt}  \citep{Turk:2010dd}.
In this way, we measure the momentum deposition rate by the inflowing gas  across the gravitational focus cross section, $\pi \Ra^2$.
 In our simulations, performed in the frame of the accretor, we observe a slowing and pileup of background gas due to gravitational interaction with the point mass. In the frame of the fluid, this represents a continuous decrease in relative velocity of the accretor, or a gravitational drag force \citep{1999ApJ...513..252O}.\footnote{This quantity can be differentiated from the {\it aerodynamic} or sometimes called {\it hydrodynamic} drag generated by a non-gravitating sphere's geometric cross section. In many cases, the geometric cross section $\pi R_{\rm obj}^2 \ll \pi \Ra^2$ (See Figure \ref{fig:scalesSummary}) and the drag generated through gravitational convergence of the background flow dominates \citep{1999ApJ...513..252O,Passy:2012jo}. If the evaluating surface in Equation \eqref{drageq} were similar to the object's size, we could instead measure the aerodynamic drag of the sink through the gas as is done by \citet{Ricker:2007kc,Ricker:2012gu}.  }
Integrated along the orbital path, this drag force produces the CE inspiral.  Thus the rate of orbital energy decay due to drag should be $\dot E_{\rm d} \approx F_{{\rm d},x} \vinf$, under the assumption that it takes many flow crossing times to dissipate the accretor's kinetic energy, $E/\dot E_{\rm d} \gg \Ra/\vinf$.

\begin{figure}[tbp]
\begin{center}
\includegraphics[width=0.45\textwidth]{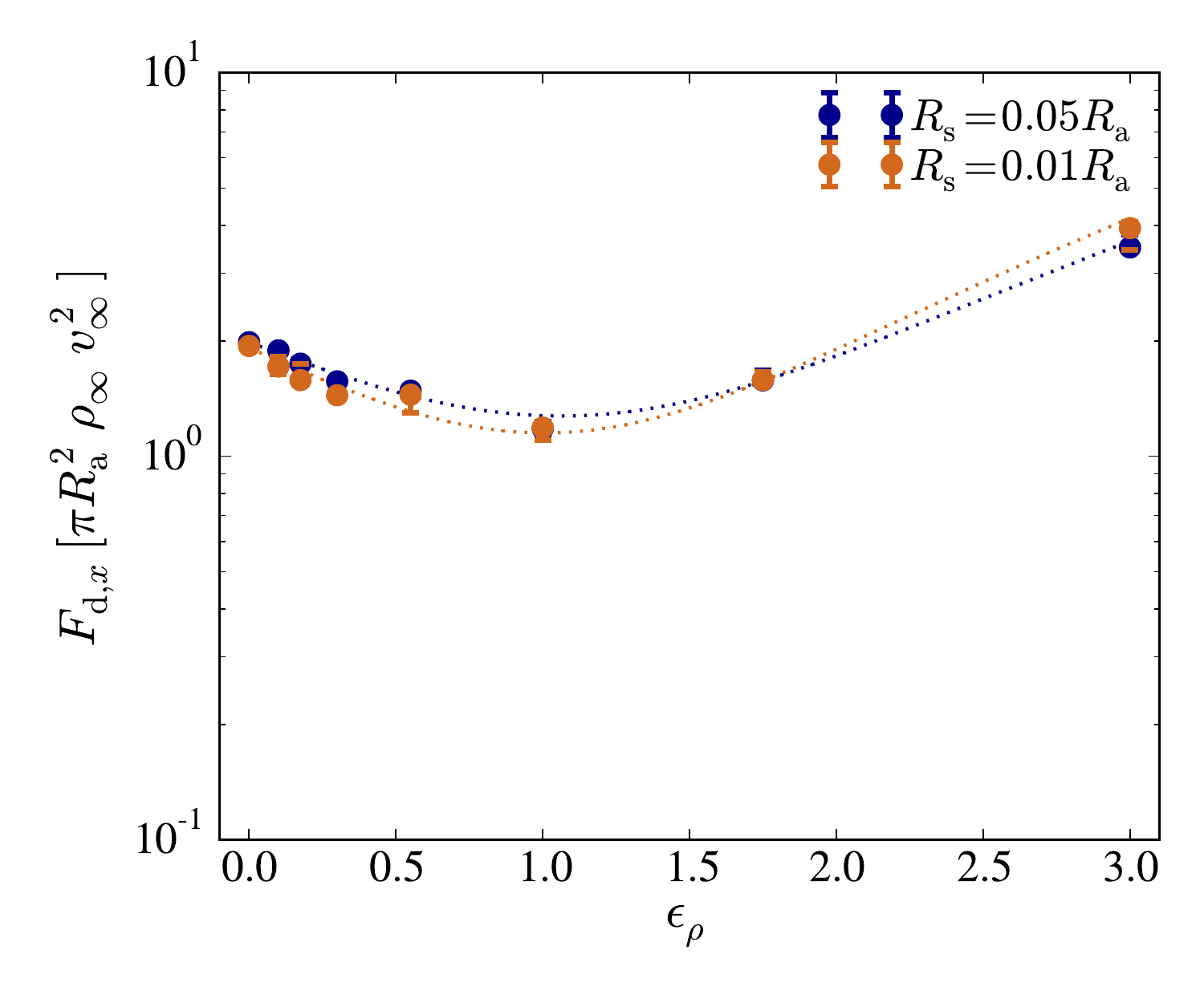}
\caption{Drag force along the direction of motion, equation \eqref{drageq}  in simulations A-H and I-P.  Drag is most in the steepest-gradient simulated, $\erho=3$. In the mild-gradient simulations, $\erho=0.1-1$, there is less drag than in the no-gradient, $\erho=0$ case. The reasons for this can be seen visually in Figure \ref{fig:sim_grad_dens}, where a trade-off can be seen in the degree of thermalization realized in the orbital plane ($x$-$y$) and in the perpendicular plane ($x$-$z$) as a gradient is introduced.  The drag realized is consistent to within an order of magnitude for each of the models, despite the accretion rate found in the simulations varying by a factor of 100.  }
\label{fig:drag}
\end{center}
\end{figure}

We find that to within a factor of $\sim 4$, the drag is similar with changing density gradient. We plot the drag rates realized in simulations A-P in Figure \ref{fig:drag}. We normalize our results to the drag expected in HLA theory, $\pi \Ra^2 \rhoinf \vinf^2$, for which the corresponding energy dissipation rate is $\dot E_{\rm HL} =  \pi \Ra^2 \rhoinf \vinf^3$.
In simulations with a mild gradient ($\erho \lesssim 1$) we find slightly lower drag than the $\erho=0$ case. This decrease can be attributed to a trade-off between stronger shocking in the $x$-$y$ plane, but weaker in the $x$-$z$ plane with the introduction of a density gradient, as can be seen visually in Figure \ref{fig:sim_grad_dens}. As the gradient steepens to $\erho=3$, the focusing of dense material in the $x$-$y$ plane dominates and the drag force again increases with respect to the nominal value. 

Our simulations are not perfectly suited to measure the drag during a CE episode. The construction of our domain is best suited to study the flow within the accretion radius. At larger distances from the accretor, the approximations made here are less valid. A primary caveat is the missing gravitational vector of the other stellar core $\vec g$.
Another consideration is that our simulation geometry is planar. Especially when $\Ra \sim R_*$, the curved geometry of the CE may depart substantially from the planar approximation. 
Both of these effects will likely play a role in shaping the shock morphology at distances $\sim \Ra$ from the accretor. Since this is where much of the thermalization occurs, we expect that some differences would arise in the full CE geometry.  
 Further, \citet{Ricker:2007kc,Ricker:2012gu} have noted that at late times in relatively equal mass CE interactions, the envelope becomes distorted well out of its hydrostatic configuration and the drag force cannot be easily related to the initial envelope properties at a given radius.
We intend to devote future simulations to study this question.

It is striking that although the accretion rate changes by nearly two orders of magnitude as the density gradient steepens, the drag changes very little. While this finding is initially surprising, its interpretation can be traced to \citet{Dodd:1952uc}'s insightful analysis of gravitational capture from a medium containing a density gradient. While the functional form and normalization of drag and accretion rates differ substantially from their derived values, their analysis did point out that drag occurs when material is focussed within the vicinity of the accretor, for example, with impact parameter $\lesssim \Ra$. To accrete, the gas must also be liberated of its angular momentum. With a stiff equation of state like $\gamma = 5/3$, disks do not form and most material is swept away from the vicinity of the accretor before it has the chance to redistribute its angular momentum. 

\section{Discussion}\label{sec:discussion}

We have shown that the morphology of flows surrounding objects embedded within a CE can be characterized by a few key parameters. Among these are the flow Mach number $\mach$, and the relative sizes of the object radius, accretion radius, and stellar envelope radius. For the cases we consider here, $R_{\rm obj}<\Ra<R_*$, so gravitational focusing is important, and the whole accretion structure is  embedded inside the CE. We've emphasized that another key flow parameter is the upstream density inhomogeneity, which can be characterized by the local ratio of accretion radius to density scale height, $\erho$. The hydrodynamical implications of these density gradients have been explored in the previous sections. 

Under the simplifying assumption that local flows inside the CE can be described by a few key parameters, the lessons from our dimensionless calculations can be applied to any relevant CE system. In this section, we highlight some considerations in extending the results of our calculations to better understand flow properties across the range of typical encounters described in Section \ref{sec:typicalencounters}.

\subsection{Cooling, Accretion, and Feedback from Embedded Objects}\label{sec:feedback}

Accretion flows toward an embedded object can only be integrated into the object if there is an effective cooling channel or if the embedded object is a BH. 
In the case of objects with a surface, accretion liberates gravitational potential energy and generates feedback -- which, if sufficiently large, can impinge on the flow field.  In the very optically thick environment of the CE photons cannot easily propagate, and feedback from accretion will be primarily mechanical rather than photoionization-driven   \citep[for example, as modeled in HLA flows by][]{2013ApJ...767..163P}. 
We have adopted a completely absorbing central boundary condition with radius in $\Rs$ in our simulations. This is useful, of course, in evaluating accretion rates for cases which can accrete, but may not be appropriate in all scenarios. To parameterize the effect that a hard, reflecting, central boundary would have on these flow morphologies we  compare the drag luminosity, $\dot E_{\rm d} \approx F_{{\rm d},x} \vinf$, to the accretion luminosity, $\dot E_{\rm a} = G M \dot M / \Rs$. 
Whether the drag luminosity or accretion luminosity is energetically dominant thus depends on both the compactness of the accretor, $\Rs$, and the accretion rate, $\dot M$.

\begin{figure}[tbp]
\begin{center}
\includegraphics[width=0.47\textwidth]{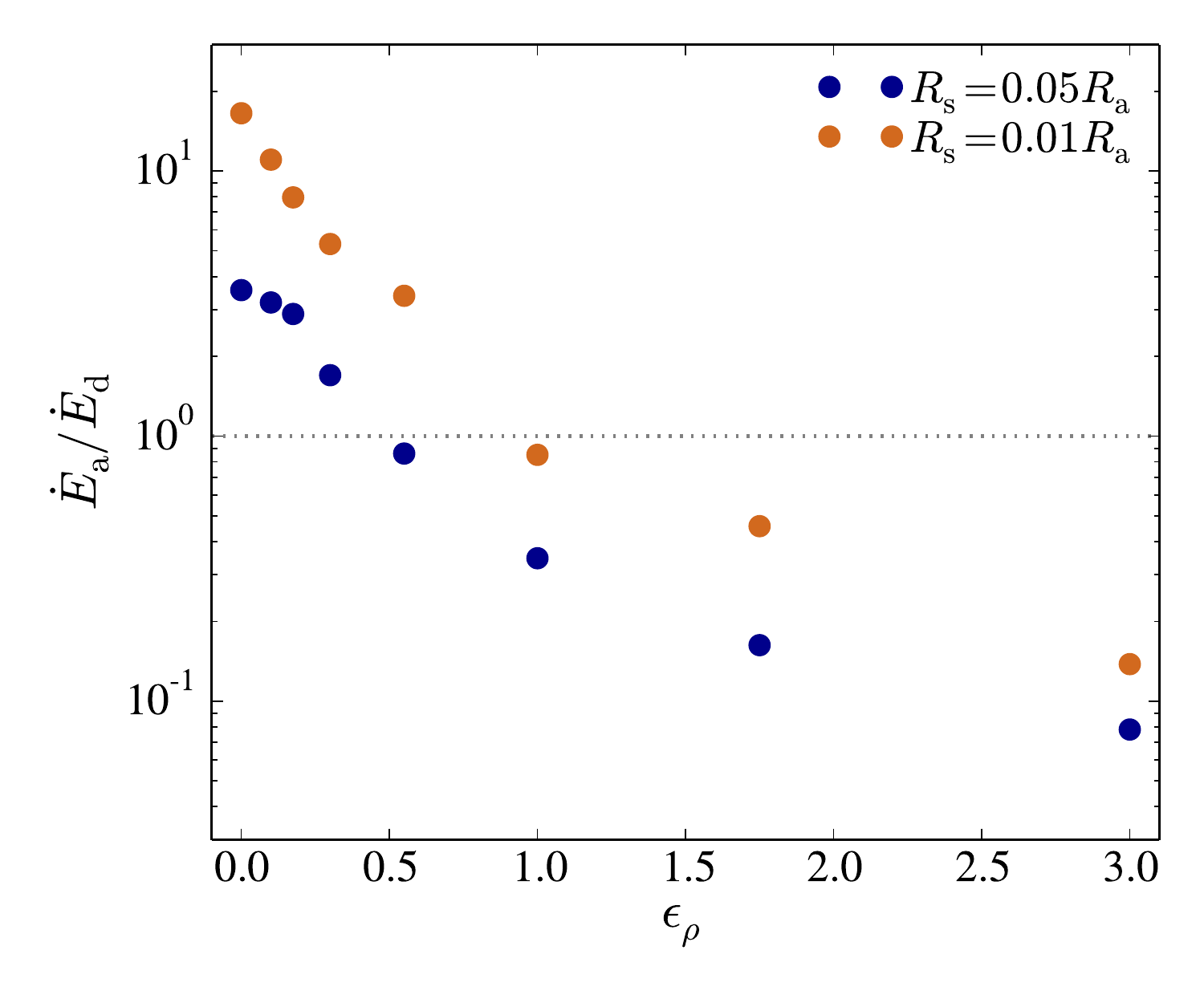}
\caption{The ratio of accretion luminosity to drag luminosity as a function of density gradient  in simulations A-H and I-P. The accretion and drag luminosity are approximated as $\dot E_{\rm} = G M \dot M / \Rs$ and $\dot E_{\rm d} = F_{{\rm d},x} \vinf $, respectively. For shallow density gradients, $\erho \lesssim 1$, feedback of accretion energy into the flow would be energetically important were one to exchange the absorbing central boundary condition for a reflecting one. For steeper gradients,  $\erho \gtrsim 1$, $\dot M$ is sufficiently small that accretion luminosity should represent only a perturbation to the energetics of the post-bow shock region.   }
\label{fig:feedback}
\end{center}
\end{figure}

Figure \ref{fig:feedback} illustrates how the ratio between accreted energy and drag-generated heat change with density gradient and sink boundary size. For regimes in which $\dot E_{\rm a}/\dot E_{\rm d} > 1$, we would expect that replacing the absorbing central boundary with a hard boundary would make a significant energetic contribution to the post-shock region. The central sink boundary acts as a cooling term in absorbing $\dot E_{\rm a}$ from the central regions. If this energy were not deleted, we might expect it to contribute to overturning the flow or modifying either the bow shock configuration or stability. On the other hand, when, $\dot E_{\rm a}/\dot E_{\rm d} < 1$ the accretion energy is small compared to the drag-generated energy in the post shock region. In this regime, the nature of the central boundary should not drastically affect the flow morphology. 
Figure \ref{fig:feedback} shows a transition between these regimes as the density gradient steepens. Steeper density gradients imply lower accretion rates into the central boundary. Because only a small fraction of material reaches the accretor, the accretion energy becomes a small contribution to the total energy budget. 

Authors who have modeled the effects of a hard central boundary condition in flow have found that feedback from the central object impinges to create a broader, more unstable bow shock \citep{Fryxell:1987eq,Zarinelli:1995wn}. It may contribute to a higher rate of vortex-shedding, and thus hydrodynamic drag \citep{Zarinelli:1995wn}, and almost certainly gives rise to a higher level of pressure support if there is angular momentum \citep{Armitage:2000eg}.
Much of this work has been performed under the assumption of 3D axisymmetry  \citep{Fryxell:1987eq} or in 2D planar geometry \citep{Zarinelli:1995wn,Armitage:2000eg}, leaving open for future investigation the details of how non-axisymmetric 3D flows respond to feedback from a hard central boundary condition and how this affects critical properties like the drag coefficient.

Extending the lessons learned in these simulations to the reality of a CE episode as it plays out in nature is admittedly somewhat more complicated. The first fact to acknowledge is that, for objects with a surface, accretion is only possible when an adequate cooling channel exists. Comparing to the examples of Section \ref{sec:typicalencounters}, the microphysics of accretion flows onto NSs is such that neutrinos can carry away the accretion energy without interacting significantly with the surrounding gas \citep{Houck:1991kc,Chevalier:1993by,Chevalier:1996kr,Fryer:1996kr}. 
Other stellar objects like main sequence stars and WDs are not sufficiently compact to promote neutrino emission, and yet the surrounding flow is extremely optically thick preventing the escape of heat through photon diffusion.  These objects, therefore, are most appropriately modeled by a hard-surface boundary condition, while NSs and BHs are more appropriately modeled with an absorbing boundary. 

Embedded objects like main sequence stars have radii of order $R_{\rm obj} \approx 0.01\Ra$ (see Table \ref{scalestable}), their inability to accrete, therefore, should affect flows with density gradients more shallow than $\erho\lesssim 1$. Embedded WDs, by contrast, have much smaller radii, $R_{\rm obj} \approx 10^{-4} \Ra$. Material that reaches these small scales will pile up rather than accreting and may contribute a substantial heating effect on the surrounding material. We cannot expect, though, to directly extrapolate our simulation $\dot M$ for $\Rs=0.01\Ra$ to $\Rs=10^{-4}\Ra$ because the flow's angular momentum as it passes the $\Rs=0.01\Ra$ indicates that only a fraction will reach the WD surface at  $10^{-4}\Ra$.

\subsection{Mass Accumulation during CE Inspiral}

NSs and BHs can gain mass by accretion during a CE episode. While the drag force, and resulting orbital energy dissipation rate $\dot E_{\rm d}$ sets the rate of inspiral, $\dot M$ sets the rate of mass growth. We can therefore compare the inspiral timescale, $E/\dot E_{\rm d}$, to the mass growth timescale $M / \dot M$. Several authors have noted that in HLA theory, the similarity in the function form of $\mhl$ and $\dot E_{\rm HL}$ imply a direct correlation between the energy required to liberate the CE and the mass accreted during the inspiral \citep{Chevalier:1993by,Brown:1995jj,Bethe:1998jv}. We can write $\dot E_{\rm HL} = \mhl \vinf^2$. Within this framework, an integrated energy injection into the CE implies an accumulated mass. In the case of a NS inspiralling through a massive companion's envelope, the implied accumulated mass would be enough to force the an accretion induced collapse to a BH. 

Recent work by \citet{Ricker:2012gu} has indicated that accretion rates from the CE might be substantially lower than the HLA value, $\mhl$. Our local simulations are complimentary to \citet{Ricker:2012gu}'s global models as we are able to allocate high resolution at scales smaller than $\Ra$ and use an absorbing sink boundary condition. We confirm that in the presence of a density gradient the accretion rate may be severely limited. We show that the accretion rate drops off drastically with increasingly steep density gradients, reaching values $\sim 10^{-2} \mhl$ for $\erho =3$. Despite this, the drag force only changes mildly in response to the density gradient. Our results therefore indicate that the mass growth timescale may become substantially longer than expected in HLA theory and that embedded objects may grow significantly less than  previously expected during CE evolution.  
We use the local calculations of accretion and drag rates presented in this paper to expand on the NS case study further in a companion paper \citep{2015ApJ...798L..19M}.

Accretion of material carrying angular momentum will impart net spin to the accreting object. If the object has radius equal to $\Rs$ used in our simulations, then $\dot L_{z}$ (Figure \ref{fig:accsummary}) can be appropriately applied. However, if $R_{\rm obj} < \Rs$, as is the case for NSs and BHs, the a more realistic approximation of $\dot L$ may be to multiply $\dot M$ by the Keplerian specific angular momentum at the object's surface: $\sqrt{G M R_{\rm obj} }$. As mentioned earlier, the extrapolation of $\dot M$ to smaller radii is not trivial. However, the fact that we find $\dot M \ll \mhl$ implies that both $\Delta M / M$ and $\Delta L / L$ are reduced proportionately with $\dot M/\mhl$.

\subsection{Loss of CE Symmetry during Inspiral}\label{sec:disturb}

In computing local simulations of CE flows, we implicitly assume minimal disturbance of the envelope during the dynamical inspiral.  While this is an extremely useful simplifying assumption, it may not be always justified in CE events as they occur in nature. The loss of spherical symmetry of the envelope material surrounding the embedded object can be crudely estimated by comparing the ratio of inspiral timescale to orbital period.  \citet{Livio:1988io} define this ratio as $\beta_{\rm CE} = (E_{\rm orb}/\dot E_{\rm d})/P_{\rm orb}$. When $\beta_{\rm CE}$ is small, local, rather than orbit-averaged effects are important. As such, large departures from the initial hydrostatic structure can be expected. Making use of the Keplerian orbital energy and period, one derives  the expression of \citet[][Equation 5]{Livio:1988io},
\beq\label{betace}
\beta_{\rm CE} \approx {1 \over 12 \pi } \left( \frac{m_*(a) + M}{M} \right) \left(\bar \rho \over \rho \right),
\eeq
where we have simplified the original author's expression by neglecting any dependence on the sound speed in the flow accretion radius $\Ra$. In the above expression the mean density is that enclosed by the orbit, $\bar \rho = m_*(a) / (4/3 \pi a^3)$.  

\begin{figure}[tbp]
\begin{center}
\includegraphics[width=0.49\textwidth]{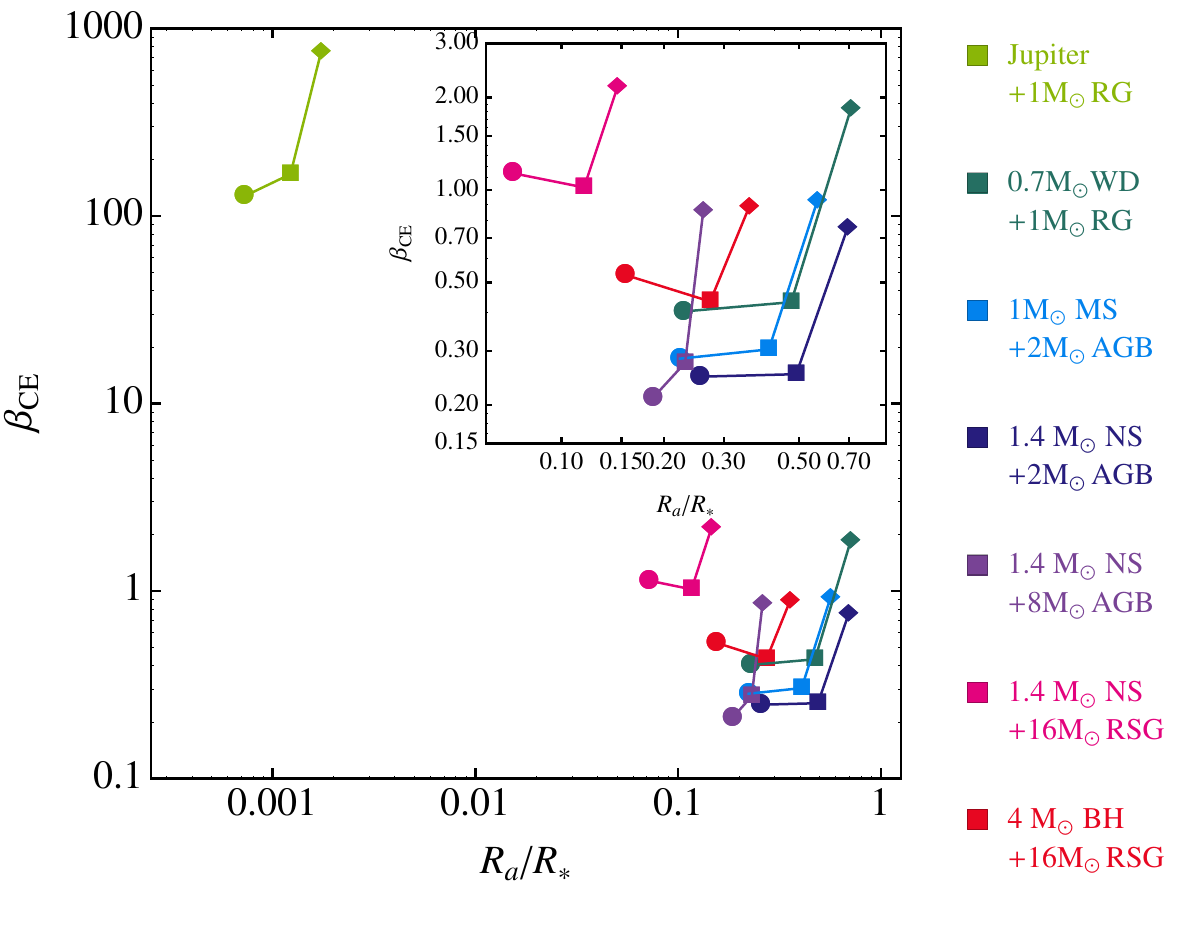}
\caption{ Characteristic scales that describe the disturbance of the CE by the embedded body. $\beta_{\rm CE}$ describes the rapidity of inspiral as compared to orbital period. Local effects should be very strong for low   $\beta_{\rm CE}$. The ratio of $\Ra/R_*$ indicates the fraction of stellar material focused to interact with the embedded body each orbit, and perhaps an indication of the shock heating. Points are evaluated at $a=$0.2, 0.5, and 0.8$R_*$ (circles, squares, and diamonds, respectively) for the same binary combinations plotted in Figures \ref{fig:scalesCompare} and \ref{fig:scalesSummary}. Our localized assumptions for CE flow properties are most valid in the upper left of this diagram, which arises when the embedded body's mass is small compared the giant-star's mass.  Among the stellar cases, a NS embedded in a supergiant companion is particularly well described by local approximation. In other cases, our local assumptions are best justified early in the CE episode where the embedded object interacts with low-density material near the stellar limb.    }
\label{fig:betaCE}
\end{center}
\end{figure}

Our local approximation of envelope properties as maintaining a quasi-hydrostatic structure similar to that of the original star is thus most justified when the embedded mass $M$ is small compared to $m_*(a)$ or when the local density $\rho \ll \bar \rho$, as is the case for highly evolved stars that develop tenuous convective envelopes. 
In Figure \ref{fig:betaCE} we compare $\beta_{\rm CE}$ to the ratio of $\Ra/R_*$ for the binary systems described in Section \ref{sec:scales}. The ratio $\Ra/R_*$ is representative of the fraction of stellar material that is being shocked in a given passage of the embedded body. The least disturbed envelopes lie at high  $\beta_{\rm CE}$ and low $\Ra/R_*$ (the upper left of the diagram).  An embedded planet is least disturbing of the examples shown, however, a NS embedded in a supergiant companion also appears to lie in the phase space best described by local approximation. 
Internal structure of the CE also plays a role. Near the stellar limb (diamond-shape points at $0.8 R_*$) where the local density is low, the local approximation is better justified than deeper in the stellar interior. 

These scalings provide some guidance in comparing our local calculations to the global calculations of \citet{Ricker:2007kc,Ricker:2012gu}  and  \citet{Passy:2012jo}. In particular, \citet{Ricker:2007kc,Ricker:2012gu} consider a 1.05$M_\odot$ red giant companion to a 0.6$M_\odot$ star. \citet{Passy:2012jo} consider a 0.88$M_\odot$ giant companion paired with point masses ranging from 0.1-0.9$M_\odot$, however, slices of the fluid conditions are only shown for the case with a 0.6$M_\odot$ point mass. Thus, these simulations, with mass ratios of order unity, lie strongly in the regime where local effects (as parameterized by $\beta_{\rm CE}$) should be important and the CE should quickly lose spherical symmetry and depart from its original structure, resulting in very rapid initial inspiral.
However, during this short phase,  flow properties compare favorably between local and global approaches. 
In particular, fluid is seen to trace out elliptical orbits around the embedded star in our local calculations, this is also observed in Figure 1 of \citet{Ricker:2007kc}. 
The bow shock structures that originate near the embedded bodies in both local and global calculations  sweep throughout the envelope to seed the the dominant spiral shock features seen in the orbital plane of global calculations \citep[e.g.][Figure 10]{Ricker:2012gu}.

\subsection{The End of Dynamical Inspiral: Envelope Spin-Up and Heating}

In CE evolution, the rapid inspiral phase precedes a gradual stabilizing of the orbital separation as orbital evolution slows to the thermal rather than dynamical timescale of the envelope \citep{Podsiadlowski:2001tn}. The rapid inspiral phase is hydrodynamic in nature, and is the phase principally addressed in simulations here as well as most recently by \citet{Passy:2012jo} and \citet{Ricker:2012gu}. During the rapid inspiral, drag luminosity, $\dot E_{\rm d}$,  transforms the orbital energy into envelope thermal energy. Drag torques due to asymmetries in the flow transfer orbital angular momentum to CE angular momentum  and orbital angular momentum is also carried away by any ejected material. Over time, the remaining CE spins closer to corotation with the embedded object \citep{Iben:1993ka}. The combination of these effects lead to lower densities, higher local sound speeds, and a progressive reduction in the flow Mach number. These effects, in turn, bring a reduction in drag, gradually ending the rapid inspiral phase. This phase of stabilization of orbital separation is observed clearly in the global simulations of \citet{Passy:2012jo} and \citet{Ricker:2012gu}. 

\begin{figure}[tbp]
\begin{center}
\includegraphics[width=0.49\textwidth]{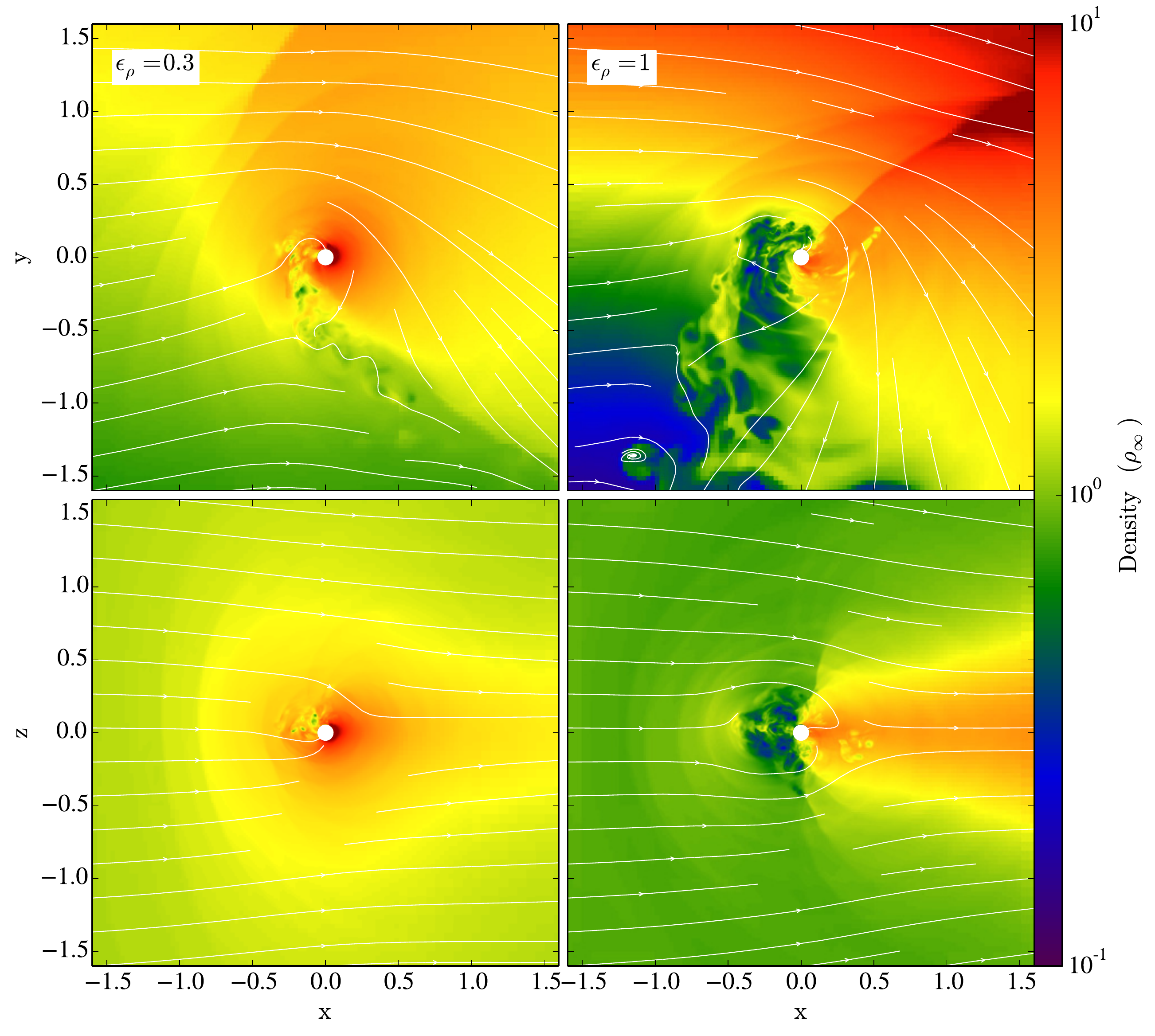}
\caption{$\mach =1.1$ simulations  T and U. These are representative of the flow at later times in the CE evolution after the envelope has been heated spun closer to corotation with the inspiralling object, and may otherwise be compared to Figure \ref{fig:sim_grad_dens}. These panels show the broader opening angle, smaller degree of tilt, and weaker shocks that develop in lower-Mach number flows. Note that the density scale here extends to $10^1$ while in Figure \ref{fig:sim_grad_dens}, it extends to $10^2$, and that the width of the panels is $3.2\Ra$ rather than $2.2\Ra$ to accommodate the larger bow-shock structures.}
\label{fig:lowmach}
\end{center}
\end{figure}

To explore qualitatively how flow morphologies might respond to reducing Mach numbers, we present two simulations in Figure \ref{fig:lowmach}. At lower Mach numbers, bow-shock features are broader, with larger opening-angles. While Figure \ref{fig:lowmach} can be compared directly to Figure \ref{fig:sim_grad_dens}, the panel width is enlarged to accommodate the larger shock-features and greater upstream standoff distance. Additionally, the colorbar shows density only up to 10 $\rhoinf$ -- representative of the fact that shocks are weaker and compression is not as severe in the $\mach=1.1$ simulations as with $\mach=2$. We also find a milder tilt-angle in the bow shock structures than in the higher Mach number simulations. As expected, the weaker shocks in this scenario lead to somewhat lower drag rates with respect to the nominal value of $\pi \Ra^2 \rho \vinf^2$ (see Table \ref{simstable}). Despite these morphological differences, we find very similar median $\dot M$ for these simulations and their $\mach=2$  counterparts.

These facts indicate that a dramatic change in flow morphology, accretion rate, and inspiral rate, probably occurs only when the the flow becomes subsonic \citep{Ricker:2012gu}. 
A final consideration on flow properties toward the end of the dynamical phase is a the decoupling of the companion's core from the envelope. In particular, when the enclosed envelope mass becomes small as compared to the binary mass that is formed by the embedded object and the core of the companion, both dense cores develop differential motion relative to the envelope. \citet{Ricker:2012gu} show that a low density, high pressure, region starts to open up surrounding the binary at this stage. Our guiding assumption of undisturbed envelope properties is clearly not realized here. Instead, typical relative motion is subsonic and local density gradients are mild. 

 In this case, the binary at the center of the extended envelope provides a heat source for all of the enshrouding CE gas.  
Modeling the subsequent thermal evolution that defines of the remainder of the CE event is certainly important in determining the outcome of the CE episode but is beyond the scope of this study.  The stabilization phase may be particularly important as the envelope may not be fully ejected at the end of the dynamical phase \citep[see][for a comparison between separations at the end of the dynamical inspiral phase and observed systems]{Passy:2012jo}. Additional terms other than orbital energy, like recombination energy, may be important to finalize the envelope ejection after the dynamical inspiral \citep{Ivanova:2014wq}.

\section{Conclusions}\label{sec:conclusion}

This paper has examined flow structures in the immediate vicinity of objects during the dynamical inspiral phase of CE. 
We begin by exploring the typical scales during the dynamical inspiral using MESA calculations of stellar structure. We then use these conditions to motivate FLASH models of accretion flows with an upstream density gradient.  We find that substantial asymmetry develops in these flows, which carry net angular momentum with respect to the embedded object. 
The main effects may be summarized as follows:
\begin{itemize}
\item Typical accretion radii, $\Ra$, are much larger than the size of embedded objects, and may in fact span a large fraction of the stellar radius. As a result, material that is focused toward the embedded object can span a large range of density as $\Ra$ sweeps across the radial density gradient within the giant star's envelope. We parameterize the density gradient according to $\erho=\Ra/\Hrho$. Typical values of $\erho$ are of order unity, so the density inhomogeneity represents a substantial perturbation to the flow. This effect is illustrated in Figure \ref{fig:densityfit} and elaborated on in Section \ref{sec:scales}.

\item  We introduce this upstream density profile into simulations of HLA. We find that the upstream gradient breaks the symmetry of the flows, because momenta no longer cancel to form an accretion column in the wake of the accretor. With an upstream gradient, the flow carries angular momentum about the accretor.  Figure \ref{fig:sim_grad_dens} illustrates the changing flow morphology with varying $\erho$. 

\item  We find that drag forces are only mildly affected by the upstream density gradient, but mass accretion rates are sensitively dependent on $\erho$. For relatively steep gradients, accretion rates can be reduced by a factor of $\sim10^{-2}$ compared to HLA theory, as shown in Figure \ref{fig:accsummary}. Because angular momentum plays a role in limiting accretion rates, we find that the accretion rates of mass and angular momentum depend on the size of the accretor, $\Rs$. We have presented simulations with $\Rs=0.01\Ra$, which is a realistic scale for a main sequence star embedded in its giant branch companion.  Drag forces, and the corresponding energy dissipation rates,  on the other hand, are only modified by a factor of a few as shown in Figure \ref{fig:drag}. 

\item Despite the presence of angular momentum in the flow, persistent disks do not form around the embedded object in our 3D simulations with a $\gamma=5/3$ equation of state. This differs from results seen in 2D \citep[for example, by][]{Armitage:2000eg}. Figure \ref{fig:disks} shows the lack of disk formation in our flow morphology, while Figure \ref{fig:steepeos} shows that a more compressible equation of state, like $\gamma=1.1$, allows the formation of a persistent disk within the bow shock. The lack of disks on large scales in CE flows may limit the degree to which viscous effects and magnetorotational instability \citep{Balbus:1991fi} can contribute to the dissipation of angular momentum and, in turn,  to accretion. 
 This lack of accretion and disk formation is consistent with the observation of \citet{2014MNRAS.439.2014T} that jets seen in post-CE planetary nebulae might most naturally arise from  interactions immediately before or after the dynamical phase.

\item In a companion paper \citep{2015ApJ...798L..19M} we apply the coefficients of drag and accretion derived here to the case of NS inspiralling through the envelope of its supergiant companion. As argued in Section \ref{sec:disturb}, this is one of the cases best described by local approximation of the CE flow properties. Due to the reduced efficiency of accretion relative to drag in the presence of a density gradient, we find that NSs undergoing typical CE episodes should accrete only a moderate mass, of order a few percent their own mass or less.  

\end{itemize}

The local calculations of flow around an embedded object during CE inspiral presented in this paper are not intended to replace global calculations. Instead, in these complementary calculations we have adopted a highly simplified description of the complex physics of a CE interaction.  Our calculations extend a tradition of numerical calculations of HLA, and treat the role of only one additional physical effect: density gradients within the CE. We find that the density gradient alone drastically modifies the flow around an embedded object as compared to homogenous HLA. This density gradient and ensuing loss of symmetry is likely responsible for the low mass accumulation rate observed by \citet{Ricker:2012gu} in their global calculation.  
Effects that may be particularly important to consider in future work include the full geometry and gravitational potential of the CE binary system as well as differential rotation and the disturbed background flow present in realistic CE events.

\begin{acknowledgements}
We are grateful for discussions with P. Armitage, L. Bildsten, C. Conroy, R. Foley, J. Forbes, C. Fryer, D. Lin, A. Loeb,  J. Naiman, R. Narayan,  P. Macias, A. MacFadyen, C. Miller,  G. Montes, D. Kasen, P. Podsiadlowski, M. Rees, S. Rosswog, M. Trenti and S. Woosley that shaped and improved this work. 
We thank the referee, Orsola De Marco, for detailed and constructive feedback.
The software used in this work was in part developed by the DOE-supported ASCI/Alliance Center for Astrophysical Thermonuclear Flashes at the University of Chicago. Simulation visualizations and analysis were made possible using the \texttt{yt} toolkit \citep{Turk:2010dd}. 
This research made use of \texttt{Astropy}, a community-developed core Python package for Astronomy \citep{2013A&A...558A..33A}.
The simulations for this research were carried out on the UCSC
supercomputer Hyades, which is supported by National Science Foundation (award number AST-1229745) and UCSC. We  acknowledge support from the David and Lucile Packard Foundation, NSF grant AST-0847563, the NSF Graduate Research Fellowship, and the Chancellor's Dissertation-Year Fellowship at UCSC. Major portions of this paper were written at DARK Cosmology Centre, Copenhagen; Aspen Center for Physics; Institute of Astronomy, Cambridge; Center for Astrophysics, Harvard and INAF-Osservatorio Astronomico di Roma.  We thank the directors of these institutions for their generous hospitality. 
\end{acknowledgements}

\bibliographystyle{apj}

\appendix
\section{Resolution Study}
\begin{figure}[tbp]
\begin{center}
\includegraphics[width=0.45\textwidth]{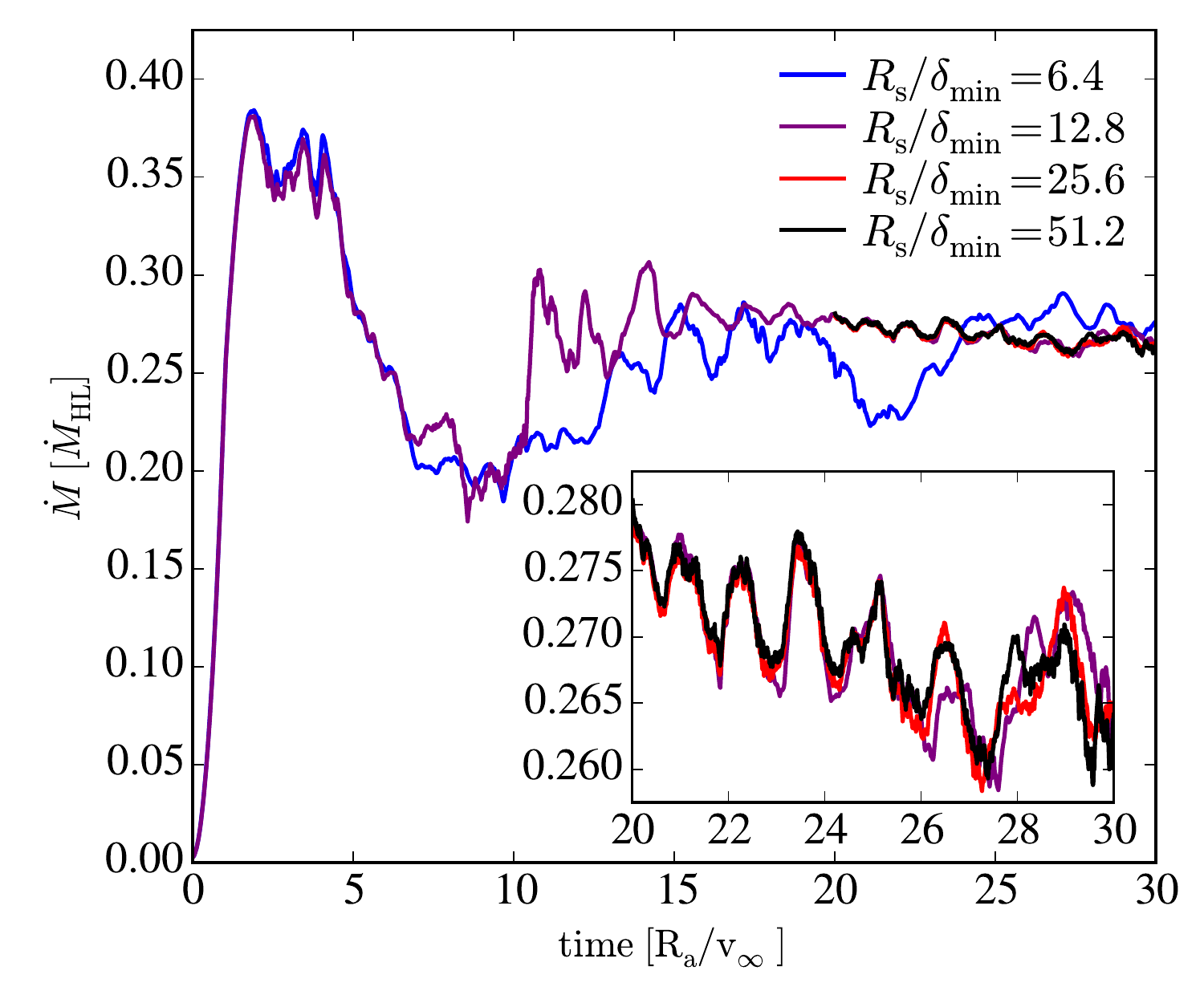}
\includegraphics[width=0.45\textwidth]{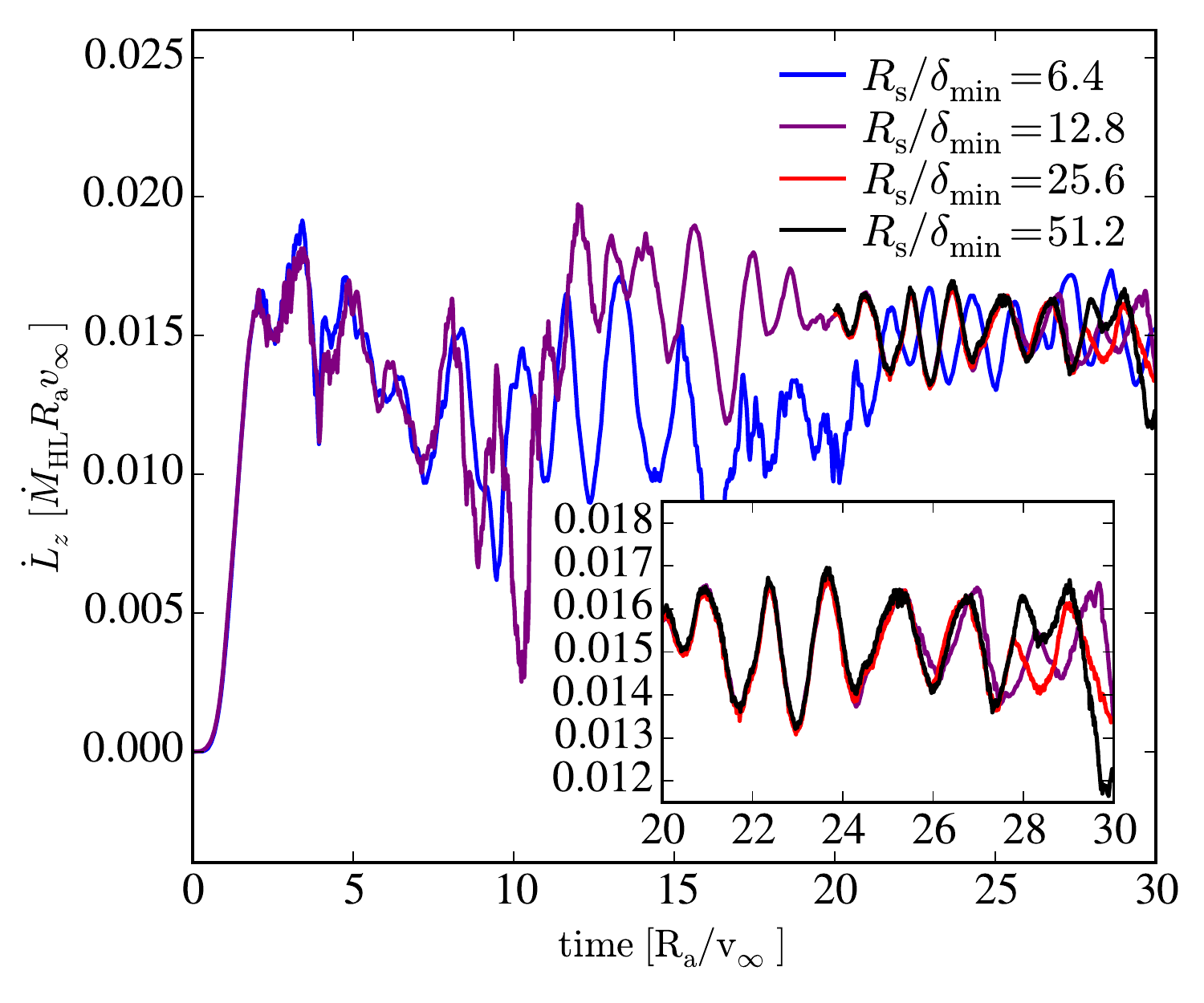}
\caption{Study of the $\erho=0.3$, $\Rs=0.05\Ra$ case with four different levels of spatial resolution  (simulations D,Q,R, and S).  The two lower level simulations were run from $t=0$, while the two higher resolution simulations were started from a checkpoint of the the $\Ra/\delta_{\rm min}=12.8$ simulation at $t=20$. There appears to be good consistency between the three highest resolution levels for 10 $\Ra/\vinf$, the crossing time of material from the box boundary. The lowest-resolution simulation exhibits larger variability than its higher-resolution counterparts.  }
\label{fig:res}
\end{center}
\end{figure}

In Figure \ref{fig:res} we confirm the robustness of our results with respect to changing spatial resolution in the $\erho=0.3$, $\Rs=0.05\Ra$ case. This comparison draws on simulations D, Q, R, and S. Because the flow is chaotic, we should not expect formal convergence, but instead aim to demonstrate that the measurement of the mass (left panel) and angular momentum (right panel) accretion rates is not impacted by our choice of numerical resolution around the accretor. We show simulations with four different levels of resolution. The three higher-resolution simulations all exhibit relatively uniform behavior in terms of mean accretion rate and amplitude of variability.  This demonstrates that numerical resolution is not significantly impacting our conclusions in terms of the total rate of accreted mass and angular momentum. 

\section{Fitting Formulae}
We derive fitting formulae for the accretion of mass, angular momentum, and for drag as a function of density gradient in our simulations. As shown in Table \ref{scalestable}, the $\Rs=0.01\Ra$ case is directly relevant to a main sequence star inside CE. Embedded compact objects like WDs, NSs, and BHs have much smaller radii and thus caution should be taken in the extrapolation of these results to those scales. 

To fit the mass accretion rate, we use a function of the form
\beq
\log (\dot M/\mhl) \approx a_1 + \frac{a_2}{1 + a_3 \erho + a_4 \erho^2}.
\eeq
We find $a_i = (-2.14034214, \ 1.94694764, \ 1.19007536,\ 1.05762477)$ for $\Rs = 0.01\Ra$ and 
$a_i = (-1.65171739, \  1.49979486, \  0.10226072, \  3.93190671)$ for $\Rs = 0.05\Ra$ via a least-squares minimization in which weights for individual points are inversely proportional to the variability defined by the 5\% and 95\% bounds of the data. The accretion of angular momentum is approximated by 
\beq
\frac{\dot L_z }{\mhl \Ra \vinf} \approx \frac{b_1 \erho }{1 + b_2 \erho + b_3 \erho^2}.
\eeq
Using the same approach, we find fitting parameters of 
$b_i = (1.86818916\times10^{-2},\  -6.42396570, \   3.40135578\times10^1)$ for $\Rs = 0.01\Ra$ and $b_i = (0.06549409, \  -6.87212261, \  27.02371844)$ for $\Rs = 0.05\Ra$.  
We fit the function 
\beq
\frac{F_{{\rm d},x} }{\pi \Ra \vinf^2} \approx c_1 + c_2 \erho + c_3 \erho^2 ,
\eeq
to the drag realized in the simulations. Fitting parameters are
$c_i = (1.91791946, \ -1.52814698, \  0.75992092)$ for $\Rs = 0.01\Ra$  and 
$c_i = (1.98255197, \ -1.33691133, \  0.62963326)$ for $\Rs = 0.05\Ra$.

\end{document}